\documentclass[aps,prd,preprintnumbers,groupedaddress,nofootinbib,amssymb,eqsecnum,notitlepage]{revtex4-1}
\usepackage{here}
\usepackage{graphicx}
\usepackage{amsmath,amsthm,amssymb}
\usepackage{bm}
\usepackage{color}

\usepackage{amsfonts}
\usepackage{dcolumn}
\allowdisplaybreaks[4]

\begin{document}
\newcommand{\newc}{\newcommand}

\newcommand{\rk}[1]{{\color{red} #1}}
\newcommand{\ben}{\begin{eqnarray}}
\newcommand{\een}{\end{eqnarray}}
\newc{\be}{\begin{equation}}
\newc{\ee}{\end{equation}}
\newc{\ba}{\begin{eqnarray}}
\newc{\ea}{\end{eqnarray}}
\newc{\bea}{\begin{eqnarray*}}
\newc{\eea}{\end{eqnarray*}}
\newc{\tp}{\dot{\phi}}
\newc{\ttp}{\ddot{\phi}}
\newc{\nrhon}{n_I\rho_{I,n_I}}
\newc{\nrhocn}{n_c\rho_{c,n_c}}
\newc{\drho}{\delta \rho_I}
\newc{\drhoc}{\delta \rho_c}
\newc{\dphi}{\delta\phi}
\newc{\D}{\partial}
\newc{\ie}{{\it i.e.} }
\newc{\eg}{{\it e.g.} }
\newc{\etc}{{\it etc.} }
\newc{\etal}{{\it et al.}}
\newcommand{\nn}{\nonumber}
\newc{\ra}{\rightarrow}
\newc{\lra}{\leftrightarrow}
\newc{\lsim}{\buildrel{<}\over{\sim}}
\newc{\gsim}{\buildrel{>}\over{\sim}}
\newc{\aP}{\alpha_{\rm P}}
\newc{\delj}{\delta j}
\newc{\rhon}{\rho_{m,n}}
\newc{\rhonn}{\rho_{m,nn}}
\newc{\delrho}{\delta \rho_m}
\newc{\pa}{\partial}
\newc{\E}{{\cal E}}
\newc{\rH}{{\rm H}}
\newc{\Mpl}{M_{\rm pl}}

\title{Weak cosmic growth in coupled dark energy 
with a Lagrangian formulation}

\author{Ryotaro Kase and Shinji Tsujikawa}

\affiliation{Department of Physics, Faculty of Science, 
Tokyo University of Science, 1-3, Kagurazaka,
Shinjuku-ku, Tokyo 162-8601, Japan}

\begin{abstract}

We investigate a dark energy scenario in which a canonical scalar 
field $\phi$ is coupled to the four velocity $u_{c}^{\mu}$ of 
cold dark matter (CDM) through a derivative interaction
$u_{c}^{\mu} \partial_{\mu} \phi$. 
The coupling is described by an interacting Lagrangian 
$f(X, Z)$, where $f$ depends on 
$X=-\partial^{\mu} \phi \partial_{\mu} \phi/2$ and 
$Z=u_{c}^{\mu} \partial_{\mu} \phi$.
We derive stability conditions of linear scalar perturbations 
for the wavelength deep inside the Hubble radius and show 
that the effective CDM sound speed is close to 0 as 
in the standard uncoupled case, while 
the scalar-field propagation speed is affected by the interacting 
term $f$. 
Under a quasi-static approximation, we also obtain
a general expression of the effective gravitational coupling 
felt by the CDM perturbation. 
We study the late-time cosmological dynamics  
for the coupling $f \propto X^{(2-m)/2}Z^m$ and show that 
the gravitational coupling weaker than the Newton constant can 
be naturally realized for $m>0$ on scales relevant to the 
growth of large-scale structures.
This allows the possibility for alleviating the tension of $\sigma_8$ 
between low- and high-redshift measurements.

\end{abstract}

\date{\today}

\pacs{04.50.Kd, 95.36.+x, 98.80.-k}

\maketitle

\section{Introduction}
\label{introsec}

The fundamental pillar in modern cosmology consists of two dark 
unknown components-- dubbed dark energy (DE) and dark matter (DM). 
The simplest candidate for DE is the cosmological constant 
$\Lambda$ \cite{Peebles:1984ge}, 
with the equation of state (EOS) corresponding to $w=-1$. 
The main source for DM is nonrelativistic 
cold dark matter (CDM) \cite{Peebles:1982ff}
with the EOS satisfying $|w| \ll 1$ to explain the observed galaxy clusterings.
The $\Lambda$CDM model is regarded as the standard cosmological paradigm, 
but it is difficult to reconcile the observed energy scale of $\Lambda$
with the vacuum energy associated with particle physics \cite{Weinberg}. 
Moreover, the tension of today's Hubble constant $H_0$ between 
low- and high-redshift measurements has been serious in the $\Lambda$CDM 
model \cite{Riess:2016jrr,Aghanim:2018eyx,Verde:2019ivm,Riess:2019cxk,Freedman:2019jwv,Wong:2019kwg,Reid:2019tiq,Shajib:2019toy}.
A similar tension is also present for the amplitude of matter density contrast 
$\sigma_8$ 
\cite{Macaulay:2013swa,Nesseris:2017vor,Hildebrandt:2016iqg,Joudaki:2017zdt} 
within the sphere of radius $8\,h^{-1}$ Mpc, 
where $H_0=100\,h$\,km\,s$^{-1}$\,Mpc$^{-1}$.

The recent studies have shown that the presence of an interaction between 
DE and DM can alleviate the aforementioned tensions of 
$H_0$ and/or $\sigma_8$ \cite{Wands,Kumar:2016zpg,Kumar:2017dnp,DiValentino:2017iww,
An:2017crg,Kaza,Yang:2018euj,Pan:2019gop,DiValentino:2019ffd,Yang:2019uog}. 
In most of these works, phenomenological interacting terms like 
$\pm \xi H \rho_{\rm DE}$ are added to the CDM and DE continuity equations 
at the background level, where $\xi$ is a coupling constant, 
$H$ is the Hubble expansion rate, and $\rho_{\rm DE}$ 
is the DE density (see Refs.~\cite{Dalal:2001dt,Zimdahl:2001ar,Chimento:2003iea,Wang1,
Wei:2006ut,Amendola:2006dg,Guo:2007zk,Gavela:2009cy,Jackson:2009mz,Faraoni:2014vra} for earlier works). 
In this approach, there is no satisfactory prescription 
for promoting the phenomenological background equations 
to their fully covariant forms. In other words, the two covariant 
theories which give rise to the same background equations can lead to 
different dynamics of cosmological perturbations. 
This pauses a problem of how to define the perturbed quantities 
properly \cite{Tamanini:2015iia}, which is related to unphysical instabilities of perturbations 
reported in Refs.~\cite{Valiviita:2008iv,Valiviita:2009nu}.

On the other hand, the Lagrangian formulation of coupled DE and DM is 
not plagued by this problem in that the interacting terms are uniquely 
fixed both at the levels of background and perturbations. 
In this vein, the theories of coupled DE and DM were constructed in 
Refs.~\cite{Pourtsidou:2013nha,Boehmer:2015kta,Boehmer:2015sha,Skordis:2015yra,
Koivisto:2015qua,Pourtsidou:2016ico,Dutta:2017kch,Kase:2019veo} by using a so-called 
Schutz-Sorkin action \cite{Sorkin,Brown} 
in the DM sector (see also Refs.~\cite{Bettoni:2011fs,Bettoni:2015wla} 
for nonminimally coupled DM). 
The perfect fluid of CDM can be described by the Schutz-Sorkin action 
containing physical quantities like the fluid density $\rho_c$ and 
the four velocity $u_c^{\mu}$ besides Lagrange multipliers.
This prescription has an advantage of dealing with both scalar and 
vector perturbations in the fluid sector on any space-time 
background \cite{DGS,DeFelice:2016yws,Kase:2018nwt,Frusciante:2018tvu,Naka2019}.

If a scalar field $\phi$ is responsible for the DE sector, 
the first possible interacting Lagrangian is of the form 
${\cal L}_{{\rm int}1}=-\sqrt{-g} f_1 (\phi, X) \rho_c (n_c)$ \cite{Boehmer:2015kta,Barros,Kase:2019veo}, 
where $g$ is the determinant of metric tensor $g_{\mu \nu}$, 
$f_1$ is a function of $\phi$ and $X=-\partial^{\mu} \phi \partial_{\mu} \phi/2$, and 
$\rho_c$ depends on the CDM number density $n_c$.
The $\phi$-dependent coupling $f_1$ arises from nonminimally 
coupled gravitational theories after the conformal transformation 
to the Einstein frame \cite{Amendola99a,Fujii}. 
In the Einstein frame, this theory corresponds to the coupled DE and DM scenario 
originally advocated in Refs.~\cite{Wetterich,Amendola99} 
(see also Refs.~\cite{Gumjudpai,Amendola06}). 
Inclusion of $X$ dependence in $f_1$ leads to different dynamics 
of background and perturbations \cite{Barros,Kase:2019veo}.

The other possible coupling between DE and CDM is a field derivative 
coupling to the fluid four velocity, which is quantified by the scalar 
combination $Z=u_c^{\mu} \partial_{\mu} \phi$. 
The interacting Lagrangian containing the linear dependence of $Z$ 
in the form ${\cal L}_{{\rm int}2}=\sqrt{-g} f_2 (n_c, \phi)Z$ 
was proposed in Ref.~\cite{Boehmer:2015sha}. 
Recently, this was further extended to 
include the $X$ dependence in $f_2$ \cite{Kase:2019veo}. 
More general coupled DE and DM theories with the nonlinear dependence 
of $Z$ were studied in Refs.~\cite{Pourtsidou:2013nha,Skordis:2015yra}. 
In Ref.~\cite{Pourtsidou:2016ico}, it was shown that the quadratic interacting 
Lagrangian of the form ${\cal L}_{{\rm int}2}=\sqrt{-g}\,Z^2$ can lead to 
an interesting possibility for alleviating the problem of $\sigma_8$ tension 
by suppressing the growth of large-scale structures. 
This property comes from a pure momentum transfer between DE and DM. 
We note that all the coupling terms mentioned above are accommodated 
by a general interacting Lagrangian of the form 
${\cal L}_{{\rm int}}=\sqrt{-g}\,f(n_c, \phi, X, Z)$, as advocated 
in Ref.~\cite{Pourtsidou:2013nha}.

In this paper, we will study the cosmological dynamics of coupled DE and DM for the 
interacting Lagrangian ${\cal L}_{{\rm int}}=\sqrt{-g}\,f(X,Z)$, where 
$f$ is a function of $X$ and $Z$. The DE and DM sectors are described by 
a nonminimally coupled scalar field with a potential $V(\phi)$ and 
a perfect fluid with the Schutz-Sorkin action, respectively. 
Unlike the analysis of 
Refs.~\cite{Boehmer:2015sha,Koivisto:2015qua,Kase:2019veo}, our interacting 
theory contains the nonlinear dependence of $Z$. 
Moreover, it can accommodate the coupling $f(Z)=Z^2$ 
mentioned above as a special case. 
It is also possible to include the dependence of $\phi$ in $f$, but 
this does not crucially modify the dynamics of perturbations. 
On the other hand, inclusion of the $n_c$ dependence in $f$ generally 
gives rise to a nonvanishing CDM sound speed \cite{Koivisto:2015qua,Kase:2019veo}, 
so we do not take it into account to avoid causing a problem for the structure formation. 

We will obtain the full linear perturbation 
equations of motion in a gauge-ready form and identify 
conditions for the absence of scalar ghosts and Laplacian instabilities. 
The effective gravitational couplings of CDM perturbations are also derived
for general functions $f(X,Z)$ under a quasi-static approximation on scales deep inside 
the Hubble radius.
We then propose a concrete model of coupled DE and DM
and study the late-time cosmological dynamics of background and 
perturbations. We show that the structure growth weaker than that 
in the $\Lambda$CDM model is naturally realized and that our coupled 
DE and DM model can ease the problem of $\sigma_8$ tension.

\section{Kinetically coupled DE and DM theories}
\label{eomsec}

We consider a canonical scalar field $\phi$ with a potential $V(\phi)$ 
derivatively coupled to the CDM with the scalar combination,  
\be
Z=u_c^{\mu} \partial_{\mu} \phi\,,
\ee
where $u_c^{\mu}$ is the CDM four 
velocity\footnote{Instead of $Z$, we can adopt the scalar combination 
$Y=J_c^{\mu} \partial_{\mu}\phi$ as in Ref.~\cite{Kase:2019veo}. 
The difference between $Y$ and $Z$ is the factor $n_c \sqrt{-g}$, 
which is constant on the cosmological background discussed in this section.}. 
The interacting action is taken to be of the form 
${\cal S}_{\rm int}=\int {\rm d}^4 x \sqrt{-g}\,f(X, Z)$, 
where $f$ is a function of 
$X=-\partial^{\mu} \phi \partial_{\mu} \phi/2$ and $Z$. 
This coupling $f$ accommodates the linear dependence 
in $Z$ \cite{Boehmer:2015sha,Koivisto:2015qua,Kase:2019veo} as well 
as the quadratic interaction $Z^2$ \cite{Pourtsidou:2016ico}. 
A more general setting with the coupling $f(n_c, \phi, X, Z)$, 
where $n_c$ is the CDM number density, was advocated in 
Ref.~\cite{Pourtsidou:2013nha}, but we consider a subclass 
of such general models for the purpose of realizing weak 
cosmic growth.
We assume that there are no direct couplings between the 
scalar field and baryons/radiations.
The gravitational sector is described by the Einstein-Hilbert action 
$S_g=\int {\rm d}^4 x  \sqrt{-g}\,(M_{\rm pl}^2/2)R$, 
where $M_{\rm pl}$ is the reduced Planck mass and $R$ 
is the Ricci scalar. The total action is then given by 
\be
{\cal S}=\int {\rm d}^4 x  \sqrt{-g} \left[ 
\frac{M_{\rm pl}^2}{2}R+X-V(\phi) \right]
-\sum_{I=c,b,r}\int {\rm d}^{4}x \left[
\sqrt{-g}\,\rho_I(n_I)
+ J_I^{\mu} \partial_{\mu} \ell_I \right]
+\int {\rm d}^4 x  \sqrt{-g}\,f(X, Z)\,.
\label{action}
\ee
The second integral corresponds to the Schutz-Sorkin action of 
perfect fluids \cite{Sorkin,Brown} describing the CDM, baryons, and radiation, 
which are labeled by $c,b,r$, respectively.
The energy density $\rho_I$ depends on the fluid number density $n_I$. 
The vector field $J_I^{\mu}$ is related to $n_I$ according to 
\be
n_I=\sqrt{\frac{J_I^{\mu} J_I^{\nu}g_{\mu \nu}}{g}}\,.
\label{nI}
\ee
The four velocity of each fluid is given by 
\be
u_{I \mu}=\frac{J_{I \mu}}{n_I \sqrt{-g}}\,,
\ee
which satisfies the relation $u_{I \mu} u_{I}^{\mu}=-1$. 
The scalar quantity $\ell_I$ is a Lagrange multiplier 
associated with the conservation of particle number, 
with the notation $\partial_{\mu } \ell_I \equiv 
\partial \ell_I/\partial x^{\mu}$.
Variation of the action (\ref{action}) with respect to  
$\ell_I$ leads to
\ba
\partial_{\mu} J^{\mu}_I=0\,,
\label{Jmu}
\ea
which holds for $I=c,b,r$. 
We note that the quantity 
$Z=J_c^{\mu} \partial_{\mu} \phi/(n_c \sqrt{-g})$ depends on 
the CDM number density through $n_c$ and 
$J_c^{\mu}$.
Varying the action (\ref{action}) 
with respect to $J_{c}^{\mu}$, it follows that 
\be
\partial_{\mu} \ell_c= \rho_{c,n_c} u_{c{\mu}}
+\frac{f_{,Z}}{n_c} \left( \partial_{\mu} \phi 
+Z u_{c{\mu}} \right)\,,
\label{lc}
\ee
where the comma in subscripts represents a partial derivative 
with respect to the scalar quantity represented in the index, 
e.g., $f_{,Z} \equiv \partial f/\partial Z$.
The corresponding relations for the baryon and radiation are
$\partial_{\mu} \ell_I= \rho_{I,n_I} u_{I{\mu}}$, with $I=b,r$. 
Later, these relations are used to eliminate the Lagrangian 
multipliers $\ell_I$ from the action (\ref{action}).

We study the cosmological dynamics of coupled DE for 
the perturbed line element given by \cite{Bardeen} 
\be
{\rm d}s^2=-(1+2\alpha) {\rm d}t^2
+2 \partial_i \chi {\rm d}t {\rm d}x^i
+a^2(t) \left[ (1+2\zeta) \delta_{ij}
+2\partial_i \partial_j E \right] {\rm d}x^i {\rm d}x^j\,,
\label{permet}
\ee
where $a(t)$ is the time-dependent scale factor.
The scalar perturbations $\alpha$, $\chi$, $\zeta$, and $E$ 
depend on both $t$ and $x^i$. 
We do not take the tensor perturbation into account in 
Eq.~(\ref{permet}), but it propagates in the same manner as in 
standard general relativity.
The scalar field $\phi$ is decomposed into the background 
part $\bar{\phi}(t)$ and the perturbation $\delta \phi$, as 
\be
\phi=\bar{\phi}(t)+\delta \phi\,,
\ee
where we will omit the bar in the following.

{}From Eq.~(\ref{Jmu}), the particle number ${\cal N}_I$ of 
each fluid is conserved 
at the background level. The temporal and spatial components of 
$J_{I}^{\mu}$ are expressed, respectively, as 
\be
J_I^0={\cal N}_I+\delta J_I\,,\qquad 
J_I^i=\frac{1}{a^2(t)} \delta^{ik} \partial_k \delta j_I\,,
\label{JI}
\ee
where $\delta J_I$ and $\delta j_I$ are 
the scalar perturbations. 
We define the velocity potentials $v_I$ according to 
\be
u_{Ii}=-\partial_{i}v_I\,.
\ee
Since $u_{Ii}=J_{Ii}/(n_I \sqrt{-g})=({\cal N}_I \partial_{i} \chi
+\partial_{i} \delta j_I)/{\cal N}_I$ for linear perturbations, 
it follows that 
\be
\partial_i \delta j_I=-{\cal N}_I \left( \partial_i \chi 
+\partial_i v_I \right)\,.
\label{delj}
\ee
On using Eq.~(\ref{lc}), there is also the following relation 
\be
\ell_c=-\int^{t} \rho_{c,n_c} (\tilde{t})\,{\rm d} \tilde{t}
+\frac{f_{,Z}}{n_c} \delta \phi
-\left( \rho_{c,n_c}+\frac{f_{,Z}}{n_c} \dot{\phi} 
\right) v_c\,, 
\label{sollc}
\ee
up to first order in perturbations. 
For the baryon and radiation, the corresponding relations 
for Lagrange multipliers are
\be
\ell_I=-\int^t\rho_{I,n_I}(\tilde{t}){\rm d}\tilde{t}
-\rho_{I,n_I}(t)v_I\,,
\label{sollb}
\ee
with $I=b,r$.

The perturbed number density of each fluid $\delta n_I$, 
which follows from Eq.~(\ref{nI}) for the 
line element (\ref{permet}), is given by 
\be
\delta n_I=\frac{1}{a^3} 
\left[ \delta J_I-{\cal N}_I \left( 3\zeta+\partial^2 E 
\right) \right]\,,
\label{delnI}
\ee
up to linear order in perturbations. 
Since the density of each fluid $\rho_I$ depends on $n_I$ alone, 
the density perturbation $\delta \rho_I$ is given by 
$\delta \rho_I=\rho_{I,n_I} \delta n_I$, so that 
\be
\delta \rho_I=\frac{\rho_{I,n_I}}{a^3} 
\left[ \delta J_I-{\cal N}_I \left( 3\zeta+\partial^2 E 
\right) \right]\,. 
\label{delrho}
\ee

At the background level, the conservation of $J_I^0$ corresponds to  
${\cal N}_I=n_I a^3={\rm constant}$. This translates to the continuity equation 
\be
\dot{\rho}_I+3H \left( \rho_I+P_I \right)=0\,,\quad {\rm for} 
\quad I=c,b,r\,,
\ee
where 
\be
P_I=n_I \rho_{I,n_I}-\rho_I
\ee
is the pressure of each fluid. 
We focus on the case in which the weak energy condition 
$\rho_I+P_I=n_I \rho_{I,n_I}>0$ is satisfied for $I=c,b,r$.
The background equations of motion can be derived by considering the time-dependent 
perturbations $\alpha(t)$, $\zeta(t)$, $\delta \phi(t)$ and expanding the 
action (\ref{action}) up to first order in these variables. 
In doing so, we eliminate the Lagrangian multipliers $\ell_I$ 
by using Eqs.~(\ref{sollc}) and (\ref{sollb}). 
This process leads to 
\ba
& & 3M_{\rm pl}^2 H^2=\rho_{\rm DE}+\rho_c+\rho_b+\rho_r\,,
\label{back1} \\
& & M_{\rm pl}^2 \left( 2\dot{H}+3H^2 \right)=
-P_{\rm DE}-P_c-P_b-P_r\,,
\label{back2} \\
& & \dot{\rho}_{\rm DE}+3H \left( \rho_{\rm DE}+P_{\rm DE} 
\right)=0\,,
\label{back3}
\ea
where $H=\dot{a}/a$ is the Hubble expansion rate, 
a dot represents the derivative with respect to $t$, and 
\ba
\rho_{\rm DE} &=& \frac{1}{2} \dot{\phi}^2+V-f
+f_{,X} \dot{\phi}^2+f_{,Z} \dot{\phi}\,,\label{rhode}\\ 
P_{\rm DE} &=& \frac{1}{2} \dot{\phi}^2-V+f\,.
\label{Pde}
\ea
The interaction between CDM and the scalar field corresponds to the momentum 
transfer \cite{Pourtsidou:2013nha,Pourtsidou:2016ico}, in which case there are 
no direct couplings appearing on the right hand sides of their continuity equations. 
More explicitly, we can write the continuity Eq.~(\ref{back3}) in the form 
\be
q_s \ddot{\phi}+3H \left( 1+f_{,X} \right) \dot{\phi}+V_{,\phi}
+3H f_{,Z}=0\,,
\label{conde}
\ee
where 
\be
q_s \equiv 1+f_{,X}+f_{,XX} \dot{\phi}^2+f_{,ZZ}
+2f_{,XZ} \dot{\phi}\,.
\label{qs}
\ee
Provided that $q_s \neq 0$, Eq.~(\ref{conde}) 
can be solved for $\ddot{\phi}$.

We define the DE equation of state $w_{\rm DE}$ 
and the effective equation of state $w_{\rm eff}$, as
\be
w_{\rm DE}=\frac{P_{\rm DE}}{\rho_{\rm DE}}\,,\qquad 
w_{\rm eff}=-1-\frac{2\dot{H}}{3H^2}\,.
\label{wdef}
\ee
As we observe in Eqs.~(\ref{back1})-(\ref{back2}) and 
Eqs.~(\ref{rhode})-(\ref{Pde}), the coupling $f$ generally 
modifies the values of $w_{\rm DE}$ and $w_{\rm eff}$ 
in standard uncoupled quintessence.

\section{Second-order scalar action and perturbation equations}
\label{persec}

We derive the full linear perturbation equations of motion for the coupled 
DE and DM theory introduced in Sec.~\ref{eomsec}. 
In the gravitational sector, we use the general perturbed 
line element (\ref{permet}) of scalar perturbations.
In the following, the notation $\varepsilon^n$ is used to describe the $n$-th 
order of perturbations. 

The perturbed number density of Eq.~(\ref{nI}), 
which is expanded up to second order, is
\be
\delta n_I= \frac{\delta \rho_I}{\rho_{I,n_I}}
-\frac{{\cal N}_I (\partial v_I)^2}{2a^5}
-(3\zeta+\partial^2E)\frac{\delta \rho_I}{\rho_{I,n_I}}
-\frac{{\cal N}_I(\zeta+\partial^2E)(3\zeta-\partial^2E)}{2a^3}
+{\cal O}(\varepsilon^3)\,.
\label{delnI2}
\ee
Substituting $\delta\rho_I$ of Eq.~(\ref{delrho}) into the 
first term on the right hand side of Eq.~(\ref{delnI2}), 
it follows that $\delta \rho_I/\rho_{I,n_I}$ is identical to 
$\delta n_I$ given by Eq.~(\ref{delnI}) at linear order in perturbations.
Introducing the fluid sound speed squared, 
\be
c_I^2=\frac{n_I \rho_{I,n_I n_I}}{\rho_{I,n_I}}\,,
\ee
the energy density $\rho_I (n_I)$ can be expanded in the form 
\be
\rho_I (n_I)=\rho_I+\left( \rho_I+P_I \right) 
\frac{\delta n_I}{n_I}+\frac{1}{2} \left( \rho_I+P_I \right) 
c_I^2 \left( \frac{\delta n_I}{n_I} \right)^2
+{\cal O} (\varepsilon^3)\,.
\ee
We also expand the interacting Lagrangian $f(X,Z)$, as
\be
f(X,Z)=f+f_{,X} \delta X+f_{,Z} \delta Z
+\frac{1}{2} f_{,XX} \delta X^2
+\frac{1}{2} f_{,ZZ} \delta Z^2
+f_{,XZ} \delta X \delta Z+{\cal O} (\varepsilon^3)\,,
\ee
where 
\ba
& &
\delta X=\dot{\phi} (\dot{\delta \phi}-\dot{\phi} \alpha)
+\frac{1}{2} \left[  (\dot{\delta \phi}-2\dot{\phi} \alpha)^2
-\frac{1}{a^2}(\partial \delta \phi+\dot{\phi} \partial \chi)^2
\right]+{\cal O} (\varepsilon^3)\,,
\label{deltaX}\\
& &
\delta Z=\dot{\delta \phi}-\dot{\phi} \alpha
+\frac{1}{2a^2} \left[ \dot{\phi} \left\{ 3a^2 \alpha^2
-(\partial_i \chi)^2+(\partial_i v_c)^2 \right\}-2a^2 \alpha \dot{\delta \phi}
-2\partial_i \delta \phi (\partial_i \chi+\partial_i v_c) 
\right] +{\cal O} (\varepsilon^3)\,.
\label{delZ}
\ea
Note that we used Eq.~(\ref{delj}) for the derivation of Eq.~(\ref{delZ}).

Expanding the action (\ref{action}) up to quadratic order in perturbations 
and using the background Eqs.~(\ref{back1})-(\ref{back3}),  
the resulting second-order action yields
\be
{\cal S}^{(2)}=\int {\rm d}t\,{\rm d}^3 x \left( L_0+L_f \right)\,,
\label{saction}
\ee
where 
\ba
\hspace{-1cm}
& &
L_0
= a^3\Biggl[ \frac{\dot{\dphi}^2}{2}-\frac{(\partial\dphi)^2}{2a^2}
-\frac{V_{,\phi \phi}}{2}\dphi^2
-\left( \dot{\phi} \dot{\delta \phi}+V_{,\phi} \delta \phi \right) \alpha
+\frac{\partial^2 \chi}{a^2} \left( \dot{\phi} \delta \phi-2H M_{\rm pl}^2 
\alpha \right)+\left( \frac{\dot{\phi}^2}{2}-3H^2 M_{\rm pl}^2 
\right)\alpha^2
\nonumber \\
\hspace{-1cm}
&&~~~~
+\sum_{I=c,b,r}\left\{ \left( \rho_I+P_I \right)v_I \frac{\partial^2 \chi}
{a^2}-v_I \dot{\delta \rho}_I-3H (1+c_I^2) v_I \delta \rho_I 
-\frac{\rho_I+P_I }{2a^2} (\partial v_I)^2
-\frac{c_I^2}{2 (\rho_I+P_I)} \delta \rho_I^2 
-\alpha \delta \rho_I \right\} 
\nonumber \\
\hspace{-1cm}
&&~~~~
+\biggl[ \frac{2M_{\rm pl}^2}{a^2} \partial^2 \chi
-3\dot{\phi} \delta \phi+6H M_{\rm pl}^2 \alpha 
-\sum_{I=c,b,r} 3(\rho_I+P_I)v_I \biggr] \dot{\zeta}
-M_{\rm pl}^2 \left[ 3 \dot{\zeta}^2+\frac{2\alpha \partial^2 \zeta}
{a^2}-\frac{(\partial \zeta)^2}{a^2} \right]
\nonumber \\
\hspace{-1cm}
&&~~~~
+\bigg\{\left[ 2M_{\rm pl}^2 \{ \ddot{\zeta}+3H \dot{\zeta}-(3H^2+\dot{H}) \alpha
-H \dot{\alpha}
\}+\dot{\phi} \dot{\delta \phi}+ \left( \ddot{\phi}+3H \dot{\phi} \right) 
\delta \phi \right]+\sum_{I=c,b,r}(\rho_I+P_I)(\dot{v}_I-3Hc_I^2v_I) 
\biggr\}
\pa^2 E \Biggr],
\label{L0}
\\
\hspace{-1cm}
& &L_f=a^3
\biggl\{ \frac{1}{2} 
\left( q_s-1 \right) \left( \dot{\phi} \alpha -\dot{\delta \phi} 
\right)^2-\frac{f_{,X}(\partial\dphi)^2}{2a^2}
-\frac{f_{,Z} \dot{\phi}(\partial v_c)^2}{2a^2}
-3 f_{,X}\dot{\phi}\dot{\zeta} \delta \phi 
+\frac{f_{,X} \dot{\phi}\,\partial^2 \chi \delta \phi}{a^2} 
-3f_{,Z} \dot{\phi} v_c \dot{\zeta}
+\frac{f_{,Z} \dot{\phi}\,v_c \partial^2 \chi}{a^2} 
\nonumber \\
\hspace{-1cm}
& &~~~~
+f_{,Z}\frac{\dot{\delta \rho}_c+3H (1+c_c^2) \delta \rho_c}
{\rho_c+P_c} \left( \delta \phi-\dot{\phi} v_c \right)
-\dot{\phi} \left( f_{,X}\,\delta \phi+f_{,Z}\,v_c \right) \partial^2 
\dot{E}
\biggr\}\,.
\label{Lf}
\ea
The Lagrangian $L_f$ arises from the coupling $f(X,Z)$.

In the following, we derive the linear perturbation equations in 
Fourier space with the comoving wavenumber $k$.
Variations of the action (\ref{saction}) with respect to nondynamical variables 
$\alpha$, $\chi$, $v_c$, $v_b$, $v_r$, and $E$ lead, respectively, to
\ba
& &
q_s \dot{\phi}\,\dot{\delta \phi}-6H M_{\rm pl}^2 \dot{\zeta}
+V_{,\phi} \delta \phi+\left( 6 H^2 M_{\rm pl}^2-q_s \dot{\phi}^2 
\right) \alpha-\frac{2k^2 M_{\rm pl}^2}{a^2} \left[ \zeta
+H \left( \chi- a^2 \dot{E} \right) \right]
+\sum_{I=c,b,r} \delta \rho_I=0\,,\label{pereq1} \\
& &
2M_{\rm pl}^2 \left( \dot{\zeta}-H \alpha \right)
+\left( 1+f_{,X} \right) \dot{\phi} \delta \phi
+\sum_{I=c,b,r} \left( \rho_I+P_I \right)v_I
+f_{,Z} \dot{\phi}\,v_c=0\,,\label{pereq2} \\
& &
\dot{\delta \rho}_I+3H \left( 1+c_I^2 \right) \delta \rho_I
+3 \left (\rho_I+P_I \right) \dot{\zeta}
+\frac{k^2}{a^2} \left( \rho_I+P_I \right) 
\left( v_I+\chi-a^2\dot{E} \right)=0\,, 
\qquad {\rm for} \quad I=c,b,r\,,\label{pereq3} \\
& &
2M_{\rm pl}^2 \left( \ddot{\zeta}+3H \dot{\zeta} \right)+
\left( 1+f_{,X} \right) \dot{\phi} \dot{\delta \phi}
+\left[ \left( 1+f_{,X}+f_{,XX} \dot{\phi}^2+f_{,XZ} \dot{\phi} 
\right) \ddot{\phi}+3H \left( 1+f_{,X} \right) \dot{\phi} \right] \delta \phi 
\nonumber \\
& &
+f_{,Z} \dot{\phi}\,\dot{v}_c
+\left( f_{,Z} \ddot{\phi}+\beta_f \dot{\phi}\right) v_c
-2M_{\rm pl}^2 \left[ (3H^2+\dot{H}) \alpha+H \dot{\alpha} \right]
+\sum_{I=c,b,r} \left( \rho_I+P_I \right) \left( \dot{v}_I -3H c_I^2 v_I 
\right)=0\,,\label{pereq4}
\label{Eeq}
\ea
where 
\be
\beta_f \equiv 
3H f_{,Z}+\left( f_{,XZ}\dot{\phi}+f_{,ZZ} \right)\ddot{\phi}\,.
\ee
Varying the action (\ref{saction}) with respect to dynamical fields 
$\delta \phi$, $\delta \rho_c$, $\delta \rho_b$, 
$\delta \rho_r$, and $\zeta$, it follows that 
\ba
& & 
\dot{\cal Y}+3H {\cal Y}+V_{,\phi} \alpha +V_{,\phi \phi} \delta \phi
+\left( 1+f_{,X} \right) \left[ 3 \dot{\phi} \dot{\zeta}+\frac{k^2}{a^2} 
\left( \dot{\phi} \chi+\delta \phi \right) \right]
-f_{,Z} \frac{\dot{\delta \rho}_c+3H (1+c_c^2)\delta \rho_c}
{\rho_c+P_c} \nonumber \\
& &+k^2 \left[ \left( 1+f_{,X}+f_{,XX} \dot{\phi}^2+f_{,XZ} \dot{\phi} 
\right) \ddot{\phi}+3\left( 1+f_{,X} \right) H \dot{\phi} \right]E=0\,,
\label{Yeq} \\
& &
\left( 1+\frac{f_{,Z}\dot{\phi}}{\rho_c+P_c} \right) 
\dot{v}_c-\left(3H c_c^2 -\frac{f_{,Z} \ddot{\phi}+\beta_f \dot{\phi}}
{\rho_c+P_c} \right) v_c-\alpha -\frac{c_c^2 \delta \rho_c+f_{,Z} \dot{\delta \phi}
+\beta_f \delta \phi}
{\rho_c+P_c}=0\,,\label{vceq} \\
& &
\dot{v}_I-3H c_I^2\,v_I-\alpha-\frac{c_I^2}{\rho_I+P_I} \delta \rho_I 
=0\,,\qquad {\rm for} \quad I=b,r\,,
\label{veq}\\
& &
\dot{\cal W}+3H {\cal W}+\sum_{I=c,b,r} (\rho_I+P_I)
(\dot{v}_I-3Hc_I^2\,v_I)+\frac{2k^2}{3a^2}M_{\rm pl}^2 
\left( \alpha+\zeta \right) 
+f_{,Z} \dot{\phi}\,\dot{v}_c+\left( f_{,Z} \ddot{\phi}+
\beta_f \dot{\phi}\right)v_c=0\,,
\label{zetaeq}
\ea
where 
\ba
& &
{\cal Y} \equiv q_s \left(  \dot{\delta \phi}-\dot{\phi}\,\alpha \right)
-k^2 \left( 1+f_{,X} \right) \dot{\phi} E\,,\\
& &
{\cal W} \equiv 2M_{\rm pl}^2 \left( \dot{\zeta}-H \alpha \right)
+\left(1+f_{,X} \right) \dot{\phi} \delta \phi
+\frac{2k^2}{3a^2}M_{\rm pl}^2 
\left( \chi-a^2 \dot{E} \right)\,.
\ea
Eliminating the second time derivative $\ddot{\zeta}$ from 
Eqs.~(\ref{Eeq}) and (\ref{zetaeq}), we obtain
\be
\Psi=-\Phi\,,
\label{aniso}
\ee
where $\Psi$ and $\Phi$ are gauge-invariant Bardeen 
gravitational potentials \cite{Bardeen} defined by 
\be
\Psi=\alpha+\frac{{\rm d}}{{\rm d}t} 
\left( \chi - a^2 \dot{E} \right)\,,\qquad 
\Phi=\zeta+H \left( \chi - a^2 \dot{E} \right)\,.
\label{PsiPhi}
\ee
{}From Eq.~(\ref{aniso}), there is no gravitational slip 
for the coupled DE and DM theory given by the action (\ref{action}). 

The perturbation equations (\ref{pereq1})-(\ref{pereq4}) and 
(\ref{Yeq})-(\ref{zetaeq}) can be applied to any gauges of interest, 
i.e., they are written in a gauge-ready 
form \cite{Hwang:2001qk,Heisenberg:2018wye}.

\section{Stability conditions and effective gravitational couplings}
\label{stasec}

In this section, we derive stability conditions for dynamical 
scalar perturbations by eliminating nondynamical variables.
We also obtain effective gravitational couplings felt by CDM and baryons 
for the perturbations relevant to the growth of large-scale structures.

\subsection{Stability conditions}

In order to discuss conditions for the absence of ghosts and Laplacian 
instabilities of scalar perturbations, we choose the unitary gauge 
characterized by 
\be
\delta \phi=0\,,\qquad E=0\,.
\label{unitary}
\ee
Under this choice, the gauge-invariant quantities,  
\be
{\cal R}=\zeta-\frac{H}{\dot{\phi}} \delta \phi\,,\qquad 
\delta \rho_{I{\rm u}}=\delta \rho_I -\frac{\dot{\rho}_I}
{\dot{\phi}}\delta \phi\,,
\ee
reduce to ${\cal R}=\zeta$ and $\delta \rho_{I{\rm u}}
=\delta \rho_I$, respectively. 

We first solve Eqs.~(\ref{pereq1})-(\ref{pereq3}) for nondynamical perturbations 
$\alpha$, $\chi$, $v_c$, $v_b$, $v_r$. 
Eliminating these quantities from Eq.~(\ref{saction}) and integrating it 
by parts, the second-order action reduces to
\be
{\cal S}^{(2)}=\int {\rm d}t\,{\rm d}^3x\,a^{3}\left( 
\dot{\vec{\mathcal{X}}}^{t}{\bm K}\dot{\vec{\mathcal{X}}}
-\frac{k^2}{a^2}\vec{\mathcal{X}}^{t}{\bm G}\vec{\mathcal{X}}
-\vec{\mathcal{X}}^{t}{\bm M}\vec{\mathcal{X}}
-\frac{k}{a}\vec{\mathcal{X}}^{t}{\bm B}\dot{\vec{\mathcal{X}}}
\right)\,,
\label{Ss2}
\ee
where ${\bm K}$, ${\bm G}$, ${\bm M}$, ${\bm B}$ 
are $4 \times 4$ matrices, with
\be
\vec{\mathcal{X}}^{t}=\left( 
{\cal R},  
\delta \rho_{c{\rm u}}/k, 
\delta \rho_{b{\rm u}}/k, 
\delta \rho_{r{\rm u}}/k \right) \,.
\label{calX}
\ee
For sufficiently small scales within the validity of 
linear perturbation theory, the nonvanishing components of 
${\bm K}$ and ${\bm G}$ are given, respectively, by 
\ba
& &
K_{11}=\frac{q_s \dot{\phi}^2}{2H^2}\,,\qquad 
K_{22}=\frac{(\rho_c+P_c+f_{,Z}\dot{\phi})a^2}{2(\rho_c+P_c)^2}
\,,\qquad 
K_{33}=\frac{a^2}{2(\rho_b+P_b)}\,,\qquad 
K_{44}=\frac{a^2}{2(\rho_r+P_r)}\,,\\
& &
G_{11}=\frac{(1+f_{,X})\dot{\phi}^2}{2H^2}\,,\qquad 
G_{22}=\frac{c_c^2\,a^2}{2(\rho_c+P_c)}\,,\qquad 
G_{33}=\frac{c_b^2\,a^2}{2(\rho_b+P_b)}\,,\qquad 
G_{44}=\frac{c_r^2\,a^2}{2(\rho_r+P_r)}\,,
\ea
where $q_s$ is defined by Eq.~(\ref{qs}). 
The leading-order components of matrix ${\bm M}$ are 
of order $k^0$. The dominant components to anti-symmetric 
matrix ${\bm B}$ are given by  
\be
B_{12}=-B_{21}=\frac{a\dot{\phi} f_{,Z}}
{2H (\rho_c+P_c)}\,,
\ee
while the other components of ${\bm B}$ are lower than the order $k^0$.

The no-ghost conditions for the curvature perturbation 
${\cal R}$ and CDM density perturbation 
$\delta \rho_{c{\rm u}}$ correspond, respectively, to
\ba
q_s &=& 
1+f_{,X}+f_{,XX} \dot{\phi}^2+f_{,ZZ}
+2f_{,XZ} \dot{\phi}>0\,,
\label{noghost1}\\
q_c &=& 1+\frac{f_{,Z}\dot{\phi}}{\rho_c+P_c}>0\,,
\label{noghost2}
\ea
where we assumed the weak energy condition 
$\rho_c+P_c>0$ for CDM.
Under the condition (\ref{noghost1}), the background field 
Eq.~(\ref{conde}) can be solved for $\ddot{\phi}$ 
without the divergence at $q_s=0$.
The $Z$ dependence in the coupling $f$ affects 
the CDM no-ghost condition (\ref{noghost2}).
Under the weak energy condition $\rho_I+P_I>0$,  
there are no ghosts for the baryon and radiation. 

The matrix components of ${\bm B}$ do not affect the dispersion 
relations for the baryon and radiation, so their 
sound speed squares $c_b^2$ and $c_r^2$ are 
equivalent to $G_{33}/K_{33}$ and $G_{44}/K_{44}$, 
respectively. For the perturbations 
${\cal X}_1 \equiv {\cal R}$ and 
${\cal X}_2 \equiv \delta \rho_{c{\rm u}}/k$,  
we first derive their equations of motion by varying 
the action (\ref{Ss2}).
Then, we substitute the solutions  
${\cal X}_j=\tilde{{\cal X}}_j e^{i (\omega t-kx)}$ 
(with $j=1, 2$) into those equations, where 
$\omega$ is a frequency.  
Since we are interested in small-scale perturbations for the 
derivation of dispersion relations, we pick up terms of 
the order $\omega^2$, $\omega k$, and $k^2$. 
This process leads to  
\ba
& & \omega^2 \tilde{{\cal X}}_1-\hat{c}_s^2\frac{k^2}{a^2}
\tilde{{\cal X}}_1-i \omega \frac{k}{a} \frac{B_{12}}{K_{11}} 
\tilde{{\cal X}}_2  \simeq 0\,,\label{dis1}\\
& & \omega^2 \tilde{{\cal X}}_2-\hat{c}_c^2\frac{k^2}{a^2}
\tilde{{\cal X}}_2-i \omega \frac{k}{a} \frac{B_{21}}{K_{22}} 
\tilde{{\cal X}}_1 \simeq 0\,,\label{dis2}
\ea
where 
\ba
\hat{c}_s^2 &=&
\frac{G_{11}}{K_{11}}=\frac{1+f_{,X}}{q_s}\,,\label{css}\\ 
\hat{c}_c^2 &=&
\frac{G_{22}}{K_{22}}=
\frac{c_c^2}{q_c}\,.
\label{csc}
\ea
In the limit that $c_c^2 \to 0$, we have $\hat{c}_c^2 \to 0$. 
In this case, the solution to Eq.~(\ref{dis2}) is given by 
\ba
& &
\omega=0\,,\label{branch1}\\
& &
\omega \tilde{\cal X}_2-i \frac{k}{a}
\frac{B_{21}}{K_{22}}
\tilde{{\cal X}}_1=0\,.\label{branch2}
\ea
The dispersion relation (\ref{branch1}) corresponds to that of CDM, 
so the CDM sound speed squared 
$c_{\rm CDM}^2=\omega^2 a^2/k^2$ is 0.
Hence the gravitational clustering of CDM density perturbations 
is not prevented or enhanced by $c_{\rm CDM}^2$.
Substituting the other solution (\ref{branch2}) into 
Eq.~(\ref{dis1}), we obtain the dispersion relation 
$\omega^2=c_s^2 k^2/a^2$, where 
\be
c_s^2=\hat{c}_s^2+\frac{f_{,Z}^2}{q_s q_c (\rho_c+P_c)}\,.
\label{csf}
\ee
The propagation speed squared $c_s^2$ of the scalar degree of freedom is 
the sum of $\hat{c}_s^2=(1+f_{,X})/q_s$ and the correction 
$f_{,Z}^2/[q_s q_c (\rho_c+P_c)]$, where the latter is 
induced by the off-diagonal components of ${\bm B}$. 
The condition for the absence of small-scale Laplacian instability 
is given by 
\be
c_s^2 \geq 0\,.
\label{cscon}
\ee
Under the no-ghost conditions (\ref{noghost1}) and (\ref{noghost2}), 
the last term in Eq.~(\ref{csf}) is positive. 
Provided that $\hat{c}_s^2 \geq 0$, the condition (\ref{cscon}) 
is always ensured. 
The above stability conditions were derived by choosing the 
unitary gauge, but they are the same for 
other gauge choices, e.g., flat  and Newtonian gauges.

\subsection{Effective gravitational couplings}

We derive the effective gravitational couplings felt by CDM and baryons 
for linear perturbations relevant to the growth of large-scale structures.
For this purpose, we consider the case in which the pressures and 
sound speed squares of both CDM and baryons vanish, i.e., 
\be
P_c=0\,,\qquad P_b=0\,,\qquad 
c_c^2=0\,,\qquad c_b^2=0\,.
\label{Pccon}
\ee
We also introduce the following gauge-invariant variables, 
\be
\delta \phi_{\rm N}=\delta \phi+\dot{\phi}\left(\chi-a^2 \dot{E}\right)\,,
\qquad 
\delta \rho_{I\rm N}=\delta \rho_I+\dot{\rho}_I \left(\chi-a^2 \dot{E}\right)\,,
\qquad 
v_{I{\rm N}}=v_I+\chi-a^2 \dot{E}\,,
\label{delphiv}
\ee
together with the gravitational potentials (\ref{PsiPhi}).
The radiation perturbation is neglected in the following discussion.
The density contrasts of CDM and baryons are given by 
\be
\delta_{I{\rm N}}\equiv \frac{\delta \rho_{I\rm N}}{\rho_I}
=\frac{\delta \rho_I}{\rho_I}
-3H \left( \chi-a^2 \dot{E} \right)\,,\quad 
{\rm with} \quad I=c,b\,.
\label{delI}
\ee
Then, we can write Eq.~(\ref{pereq3}) in the form 
\be
\dot{\delta}_{I{\rm N}}
+\frac{k^2}{a^2} v_{I{\rm N}}+3\dot{\Phi}=0\,.
\label{delIeq}
\ee
For CDM and baryons, Eqs.~(\ref{vceq}) and (\ref{veq}) give
\ba
& &
\dot{v}_{c{\rm N}}=
\frac{\rho_c \Psi+f_{,Z}(\dot{\delta \phi}_{\rm N}-\ddot{\phi} v_{c{\rm N}})
+\beta_f (\delta \phi_{\rm N}-\dot{\phi} v_{c{\rm N}})}
{\rho_c+f_{,Z} \dot{\phi}}\,,\label{vcN}\\
& &
\dot{v}_{b{\rm N}}=\Psi\,.\label{vbN}
\ea
Taking the time derivative of Eq.~(\ref{delIeq}) and 
using Eqs.~(\ref{vcN})-(\ref{vbN}), 
it follows that 
\ba
& &
\ddot{\delta}_{c{\rm N}}+\left( 2H
+\frac{f_{,Z} \ddot{\phi}+\beta_f \dot{\phi}}
{\rho_c+f_{,Z}\dot{\phi}} \right) \dot{\delta}_{c{\rm N}}
+\frac{k^2}{a^2} \frac{\rho_c \Psi+f_{,Z}\dot{\delta \phi}_{\rm N}
+\beta_f \delta \phi_{\rm N}}
{\rho_c+f_{,Z} \dot{\phi}} +3\ddot{\Phi}+3\left( 2H
+\frac{f_{,Z} \ddot{\phi}+\beta_f \dot{\phi}}{\rho_c+f_{,Z}\dot{\phi}} 
\right) \dot{\Phi} =0\,,\label{delceq} \\
& &
\ddot{\delta}_{b{\rm N}}+2H\dot{\delta}_{b{\rm N}}
+\frac{k^2}{a^2} \Psi+3\ddot{\Phi}+6H \dot{\Phi}=0\,. 
\label{delbeq}
\ea

We first express the other perturbation equations derived in Sec.~\ref{persec} 
in terms of gauge-invariant variables introduced in Eqs.~(\ref{PsiPhi}), 
(\ref{delphiv}), and (\ref{delI}). 
Then, we employ the quasi-static approximation 
for modes deep inside the sound horizon, under which the dominant contributions to 
the perturbation equations are those containing $\delta_{c{\rm N}}$, 
$\delta_{b{\rm N}}$, and $k^2/a^2$. We do not neglect the field mass 
squared $V_{,\phi \phi}$ to accommodate the case in which the field 
is heavy in the past.
Under this approximation scheme, Eqs.~(\ref{pereq1}) and (\ref{Yeq}) give
\ba
& & 
\Phi \simeq \frac{a^2(\rho_c \delta_{c{\rm N}}+
\rho_b \delta_{b{\rm N}})}{2M_{\rm pl}^2 k^2}\,,\\
& &
\delta \phi_{\rm N} \simeq \frac{a^2}{k^2} 
\frac{f_{,Z}}{1+f_{,X}+\mu_M} \dot{\delta}_{c{\rm N}}\,,
\label{delphiN}
\ea
where 
\be
\mu_M \equiv V_{,\phi \phi} \frac{a^2}{k^2}\,.
\label{muM}
\ee
On using Eq.~(\ref{aniso}), we obtain
\be
\Psi=-\Phi \simeq -\frac{a^2(\rho_c \delta_{c{\rm N}}+
\rho_b \delta_{b{\rm N}})}{2M_{\rm pl}^2 k^2}\,.
\label{Psi}
\ee

Under the quasi-static approximation, the terms containing $\ddot{\Phi}$ 
and $\dot{\Phi}$ in Eqs.~(\ref{delceq}) and (\ref{delbeq}) can be 
neglected relative to other terms.  
The $Z$ dependence in $f$ modifies the standard term 
$(k^2/a^2) \Psi$ in the left hand side of the CDM 
perturbation Eq.~(\ref{delceq}).
The effective gravitational coupling of CDM is also affected 
by the term $\dot{\delta \phi}_{\rm N}$, which contains
the second derivative $\ddot{\delta}_{c{\rm N}}$ from 
Eq.~(\ref{delphiN}).
We eliminate the terms $\dot{\delta \phi}_{\rm N}$, 
$\delta \phi_{\rm N}$, $\Psi$ in Eqs.~(\ref{delceq}) and (\ref{delbeq}) 
by using Eqs.~(\ref{delphiN}) and (\ref{Psi}).
This process leads to
\ba
&&
\ddot{\delta}_{c{\rm N}}+\left( 2+c \right)H\dot{\delta}_{c{\rm N}}
-\frac{3H^2}{2G} \left(G_{cc}\Omega_c\delta_{c{\rm N}}
+G_{cb}\Omega_b\delta_{b{\rm N}}\right)
\simeq0\,,\label{delceq2} \\
&&
\ddot{\delta}_{b{\rm N}}+2H\dot{\delta}_{b{\rm N}}
-\frac{3H^2}{2G}\left(G_{bc}\Omega_c\delta_{c{\rm N}}
+G_{bb}\Omega_b\delta_{b{\rm N}}\right)\simeq0\,, 
\label{delbeq2}
\ea
where $G=1/(8\pi M_{\rm pl}^2)$ is the Newton gravitational constant, 
and 
\be
c=\frac{(1+f_{,X}+\mu_M)^2(\beta_f \dot{\phi}+f_{,Z}\ddot{\phi})
+f_{,Z}(1+f_{,X}+\mu_M)(2\beta_f-3H f_{,Z})
-f_{,Z}^2 (f_{,XX} \dot{\phi} \ddot{\phi}+f_{,XZ} \ddot{\phi}
+\dot{\mu}_M)}
{H(1+f_{,X}+\mu_M)[(1+f_{,X}+\mu_M)(\rho_c+f_{,Z} \dot{\phi})+f_{,Z}^2]}\,.
\ee
The effective gravitational couplings for CDM and baryons are 
given, respectively, by 
\ba
& &
G_{cc}=G_{cb}=\frac{G}{1+r_f}\,,
\label{Gcc} \\
& &
G_{bc}=G_{bb}=G\,,
\label{Gbc}
\ea
where
\be
r_f \equiv \frac{f_{,Z}\,\dot{\phi}(1+f_{,X}+\mu_M)+f_{,Z}^2}
{(1+f_{,X}+\mu_M)\rho_c}\,.
\label{rf}
\ee
The $Z$ dependence in $f$ generally gives rise to the CDM gravitational 
couplings $G_{cc}$ and $G_{cb}$ different from $G$. 
If $r_f>0$, then we have $G_{cc}=G_{cb}<G$.
The baryon gravitational couplings $G_{bc}$ and $G_{bb}$ are equivalent 
to $G$, but the growth of $\delta_{b{\rm N}}$ is affected by CDM perturbations
through the density contrast $\delta_{c{\rm N}}$ in Eq.~(\ref{delbeq2}). 

For the scalar field relevant to dark energy, the field mass squared 
$V_{,\phi \phi}$ in the late Universe is typically of order $H_0^2$. 
In this case, the quantity $\mu_M$ is much smaller than 1 
for the perturbations deep inside the Hubble radius today. 
In the massless limit ($\mu_M \to 0$), we have 
\be
r_f=\frac{f_{,Z}\,\dot{\phi}(1+f_{,X})+f_{,Z}^2}
{(1+f_{,X})\rho_c}\,,
\label{rf2}
\ee
which is a key quantity for studying the evolution of matter perturbations 
at low redshifts.

\section{Concrete model}
\label{modelsec}

We propose a concrete model in a class of coupled DE and DM 
theories given by the action (\ref{action}) and study the 
cosmological dynamics of background and perturbations.
In Ref.~\cite{Pourtsidou:2013nha}, the interaction of the form 
$f=\beta Z^2$ was proposed, where $\beta$ is a constant. 
In this case, the coupling $f$ gives rise to the contribution 
$\beta \dot{\phi}^2$ to the background DE 
density (\ref{rhode}) and pressure (\ref{Pde}). 
We propose the extended version of this coupling, which is given by 
\be
f(X, Z)=\beta \left( \sqrt{2X} \right)^{2-m} Z^m\,,
\label{fun}
\ee
where $\beta$ and $m$ are constants. 
The interacting DE and DM scenario studied in 
Refs.~\cite{Pourtsidou:2013nha,Pourtsidou:2016ico} corresponds to 
the power $m=2$. 
Since $X=\dot{\phi}^2/2$ and $Z=\dot{\phi}$ at the background level, 
the coupling (\ref{fun}) reduces to $f=\beta \dot{\phi}^2$. 
The background DE density (\ref{rhode}) and 
pressure (\ref{Pde}) are given, respectively, by 
\ba
\rho_{\rm DE} &=& 
\frac{1}{2} \left(1 +2 \beta \right) \dot{\phi}^2+V\,,
\label{rhode2}\\
P_{\rm DE} &=&
\frac{1}{2} \left(1 +2 \beta \right) \dot{\phi}^2-V\,,
\label{Pde2}
\ea
which show that the modification from the interaction appears 
only through the term $\beta \dot{\phi}^2$.
Since $\rho_{\rm DE}$ and $P_{\rm DE}$ do not contain the power $m$, 
the background dynamics is independent of $m$. 
As we will see below, this is not the case for the evolution 
of cosmological perturbations. 

It is also possible to consider the more general coupling $f=\beta ( \sqrt{2X} )^{n} Z^m$, 
whose background value is $\beta \dot{\phi}^{n+m}$.
If $n+m \neq 2$, then the coupling term $\beta \dot{\phi}^{n+m}$ scales 
in a different way compared to the kinetic energy $\dot{\phi}^2/2$.
Then, the former can dominate over the latter either in the early or late
cosmological epoch. In this paper we will not study such general cases, 
but we focus on the power satisfying the condition $n+m=2$.

Let us consider CDM and baryons satisfying the conditions (\ref{Pccon}). 
For the coupling (\ref{fun}), the stability conditions (\ref{noghost1}), 
(\ref{noghost2}), and (\ref{cscon}) reduce, respectively, to 
\ba
& &
q_s=1+2\beta>0\,,
\label{qscon} \\
& &
q_c=1+\frac{\beta m \dot{\phi}^2}{\rho_c}>0\,,
\label{qccon} \\
& &
c_s^2=\hat{c}_s^2
+\frac{m^2 \beta^2 \dot{\phi}^2}
{(1+2\beta)(m \beta \dot{\phi}^2+\rho_c)}
\geq 0\,,
\label{cscon2}
\ea
where 
\be
\hat{c}_s^2=1-\frac{m \beta}{1+2\beta}\,.
\label{hatcs}
\ee
For positive values of $\beta$ and $m$, the inequalities
(\ref{qscon}) and (\ref{qccon}) trivially hold\footnote{We note that 
the sign of $\beta$ is opposite to that used in Refs.~\cite{Pourtsidou:2013nha,Pourtsidou:2016ico}.}. 
In the early matter era during which the CDM density dominates over the 
scalar-field density ($\rho_c \gg m \beta \dot{\phi}^2$) with $m \beta$ 
at most of order 1, the scalar 
sound speed squared (\ref{cscon2}) reduces to 
$c_s^2 \simeq 1-m \beta/(1+2\beta)$.
To satisfy the condition $c_s^2 \geq 0$ in this regime, we require 
that $1+\beta (2-m) \geq 0$, which is automatically satisfied 
for $0 \le m \le 2$. If $m>2$, the constant $\beta$ needs to 
be in the range $0 \le \beta \le 1/(m-2)$. 
The last term in Eq.~(\ref{cscon2}), which arises from the off-diagonal 
components of matrix ${\bm B}$ in Eq.~(\ref{Ss2}),
gives a positive contribution to $c_s^2$.
In the regime where the scalar-field density dominates over the CDM 
density ($m \beta \dot{\phi}^2 \gg \rho_c$), it follows that 
$c_s^2 \simeq 1$.

\subsection{Background dynamics}

To study the background cosmological dynamics, it is convenient to 
define the dimensionless variables,
\be
x_1=\frac{\dot{\phi}}{\sqrt{6}H M_{\rm pl}}\,,\quad 
x_2=\frac{\sqrt{V}}{\sqrt{3}H M_{\rm pl}}\,,\quad
\Omega_c=\frac{\rho_c}{3H^2 M_{\rm pl}^2}\,,\quad
\Omega_b=\frac{\rho_b}{3H^2 M_{\rm pl}^2}\,,\quad
\Omega_r=\frac{\rho_r}{3H^2 M_{\rm pl}^2}\,,\quad
\lambda=-\frac{M_{\rm pl}V_{,\phi}}{V}\,.
\ee
Then, the Hamiltonian constraint (\ref{back1}) translates to 
\be
\Omega_c=1-\left( 1+2\beta \right) x_1^2-x_2^2 
-\Omega_b-\Omega_r\,.
\ee
On using the background Eqs.~(\ref{back2}) and 
(\ref{back3}), we obtain the following dynamical equations,
\ba
x_1' &=& \frac{1}{2}x_1 \left[ 3(1+2\beta) x_1^2-3x_2^2-3+\Omega_r 
\right]+\frac{\sqrt{6}}{2(1+2\beta)} \lambda x_2^2\,,\label{dy1}\\
x_2' &=& \frac{1}{2} x_2 \left[ 3+3(1+2\beta)x_1^2 
-\sqrt{6} \lambda x_1-3x_2^2+\Omega_r \right]\,,\\
\Omega_b' &=& \Omega_b \left[ 3(1+2\beta) x_1^2 
-3x_2^2 +\Omega_r \right]\,,\\
\Omega_r' &=& \Omega_r \left[ 3(1+2\beta) x_1^2 
-3x_2^2 +\Omega_r-1 \right]\,,\label{dy4}
\ea
where a prime represents the derivative with respect to $\ln a$.
The equations of state defined in Eq.~(\ref{wdef}) reduce to 
\be
w_{\rm DE}=\frac{(1+2\beta)x_1^2-x_2^2}
{(1+2\beta)x_1^2+x_2^2}\,,\qquad
w_{\rm eff}=\left(1 +2\beta \right)x_1^2-x_2^2
+\frac{1}{3} \Omega_r\,.
\ee
The variable $\lambda$ obeys the differential equation 
\be
\lambda'=-\sqrt{6} \lambda^2 \left (\Gamma-1 
\right) x_1\,,
\ee
where $\Gamma=V V_{,\phi \phi}/V_{,\phi}^2$. 
The exponential potential 
\be
V(\phi)=V_0 e^{-\lambda \phi/M_{\rm pl}}
\label{potential}
\ee
corresponds to the case in which $\Gamma=1$. 
In this case, $\lambda$ does not vary in time, so 
the dynamical system (\ref{dy1})-(\ref{dy4}) is closed.

Let us discuss the cosmological dynamics for the exponential 
potential (\ref{potential}). 
The fixed points relevant to the radiation and matter eras 
correspond, respectively, to 
${\rm P}_r$: $(x_1, x_2, \Omega_b, \Omega_r)=(0,0,0,1)$ and 
${\rm P}_m$: $(x_1, x_2, \Omega_b, \Omega_r)=(0,0,1-\Omega_c,0)$.
The critical point responsible for cosmic acceleration  
is given by\footnote{There exists the other scaling fixed point 
$(x_1, x_2, \Omega_b, \Omega_r)=(\sqrt{6}/(2\lambda), 
\sqrt{3(1+2\beta)/(2\lambda^2)},1-\Omega_c-3(1+2\beta)/\lambda^2,0)$ with 
$w_{\rm DE}=w_{\rm eff}=0$, but this does not satisfy the condition 
for cosmic acceleration.} 
\be
{\rm P}_{\rm DE}:~(x_1, x_2, \Omega_b, \Omega_r)=
\left( \frac{\lambda}{\sqrt{6} (1+2\beta)}, 
\sqrt{1-\frac{\lambda^2}{6(1+2\beta)}},0,0 \right)\,,
\label{Pdef}
\ee
with $\Omega_c=0$, and 
\be
w_{\rm DE}=w_{\rm eff}=-1+\frac{\lambda^2}{3(1+2\beta)}\,.
\label{wac}
\ee
The cosmic acceleration occurs for $w_{\rm eff}<-1/3$, i.e., 
\be
\lambda^2<2 \left( 1+2\beta \right)\,,
\label{lamcon}
\ee
so that the positive coupling $\beta$ allows the wider range of $\lambda$ 
in comparison to the case $\beta=0$. 
For increasing $\beta$, both $w_{\rm DE}$ and $w_{\rm eff}$ 
in Eq.~(\ref{wac}) get closer to $-1$. 
Introducing a rescaled field $\varphi=\sqrt{1+2\beta}\,\phi$, 
we can express Eqs.~(\ref{rhode2}) and (\ref{Pde2}) in the forms 
$\rho_{\rm DE}=\dot{\varphi}^2/2+V_0 e^{-\tilde{\lambda} \varphi/M_{\rm pl}}$ 
and 
$P_{\rm DE}=\dot{\varphi}^2/2-V_0 e^{-\tilde{\lambda} \varphi/M_{\rm pl}}$, 
where $\tilde{\lambda} \equiv \lambda/\sqrt{1+2\beta}$. 
Hence the DE dynamics is identical to that of the canonical 
field $\varphi$ with the exponential potential $V_0 e^{-\tilde{\lambda} \varphi/M_{\rm pl}}$.
The positive coupling $\beta$ works to reduce the value of 
the effective slope $\tilde{\lambda}$ of potential.

\begin{figure}[h]
\begin{center}
\includegraphics[height=3.2in,width=3.5in]{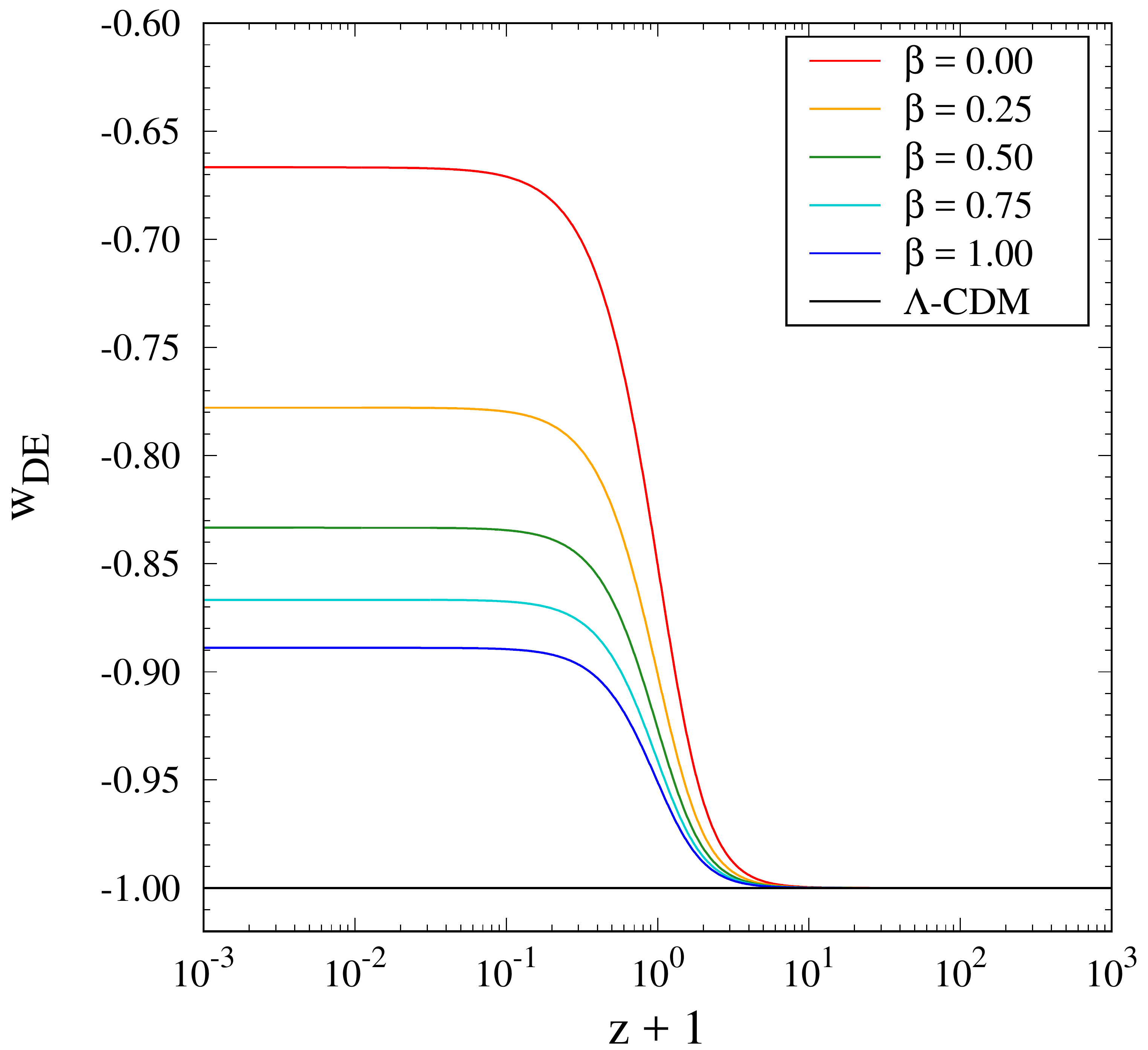}
\end{center}
\caption{\label{fig1}
Evolution of $w_{\rm DE}$ versus $z + 1$ 
for the exponential potential $V=V_0e^{-\lambda\phi/\Mpl}$ 
with $\lambda=1$ and five different values of $\beta$. 
We also plot the evolution of $w_{\rm DE}$ for the $\Lambda$CDM 
model (black line). 
The initial conditions are chosen to be $x_1=10^{-13}$, $x_2=10^{-14}$, 
$\Omega_r=0.999966$, and $\Omega_b=5.3\times10^{-6}$ around 
$z\simeq10^{8}$, so that the present epoch ($z=0$) corresponds to 
$\Omega_{\rm DE}=0.68$, $\Omega_r \simeq 10^{-4}$,  and 
$\Omega_b \simeq 0.05$.}
\end{figure}

In Fig.~\ref{fig1}, we plot the evolution of $w_{\rm DE}$ versus $z+1$ 
for $\lambda=1$ with five different values of $\beta$, 
where $z=a_0/a-1$ is the redshift with today's scale factor $a=a_0$. 
We have chosen the initial conditions $x_1 \gg x_2$ in the deep radiation 
era, in which case $w_{\rm DE}$ starts to evolve from the value away from $-1$. 
However, $w_{\rm DE}$ approaches the value close to $-1$ before the 
onset of matter era. After the dominance of DE density over the matter 
density, $w_{\rm DE}$ starts to deviate from $-1$. 
For increasing $\beta$ from 0, the deviation of $w_{\rm DE}$ from 
$-1$ at low redshifts tends to be smaller.
Thus, even for $\lambda={\cal O}(1)$, the coupling 
with $\beta>0$ allows an intriguing possibility for the better 
compatibility with observational data 
(see Refs.~\cite{Heisenberg:2018yae,Akrami} for recent observational 
constraints on uncoupled 
quintessence with the exponential potential).

%
\subsection{Evolution of perturbations}

We proceed to study the evolution of perturbations for the modes 
relevant to the growth of large-scale structures.
Since the scalar potential $V$ is at most of 
order $M_{\rm pl}^2 H^2$, the field mass squared $V_{,\phi \phi}$ 
can be estimated as $V_{,\phi \phi}=\lambda^2 V/M_{\rm pl}^2 
\lesssim \lambda^2 H^2$. 
To realize the late-time cosmic acceleration, we consider
the potential whose slope is in the range $|\lambda| \lesssim {\cal O}(1)$. 
Then, the quantity (\ref{muM}) is at most of order 
$\mu_M \lesssim (aH/k)^2$, which is much smaller than 1 
for perturbations deep inside the Hubble radius. 
Hence we set $\mu_M=0$ in the following discussion.

For the function (\ref{fun}), the gravitational couplings of CDM are 
given by $G_{cc}=G_{cb}=G/(1+r_f)$, with 
\be
r_f=\frac{2\beta m x_1^2 (1+2\beta)}
{\Omega_c [1+\beta (2-m)]}=
\frac{2\beta m x_1^2}{\Omega_c \hat{c}_s^2}\,,
\label{rfes}
\ee
where $\hat{c}_s$ is given by Eq.~(\ref{hatcs}).
Provided that the Laplacian instability of scalar 
perturbations is absent, the quantity $r_f$ is 
positive for $\beta m>0$.
When $\beta>0$ and $m>0$, the other stability conditions 
(\ref{qscon}) and (\ref{qccon}) are also automatically satisfied. 
This means that, for positive $\beta$ and $m$ in the range 
$1+\beta (2-m)>0$, both $G_{cc}$ and $G_{cb}$ are 
smaller than $G$.

\begin{figure}[h]
\begin{center}
\includegraphics[height=3.2in,width=3.5in]{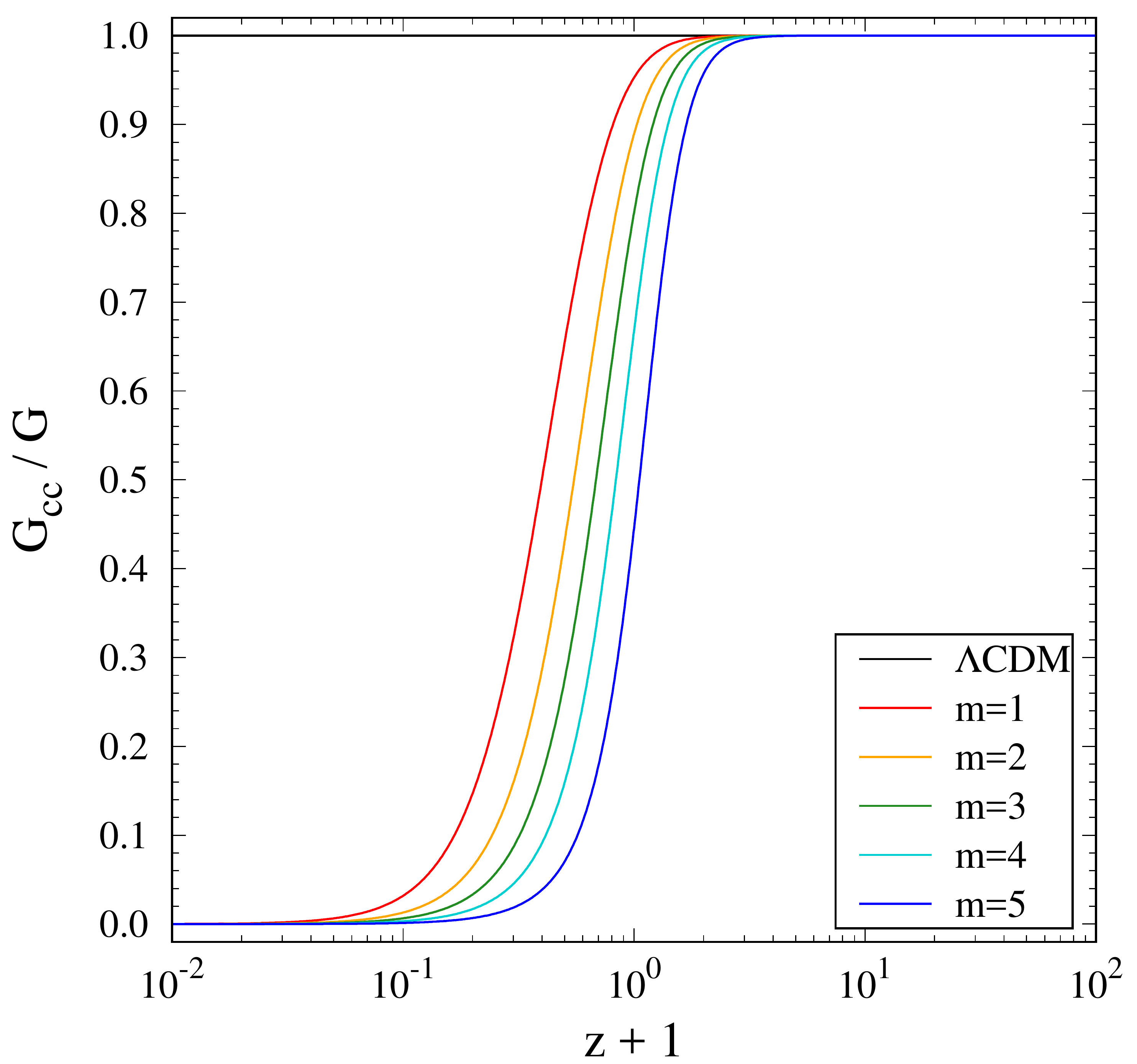}
\includegraphics[height=3.2in,width=3.5in]{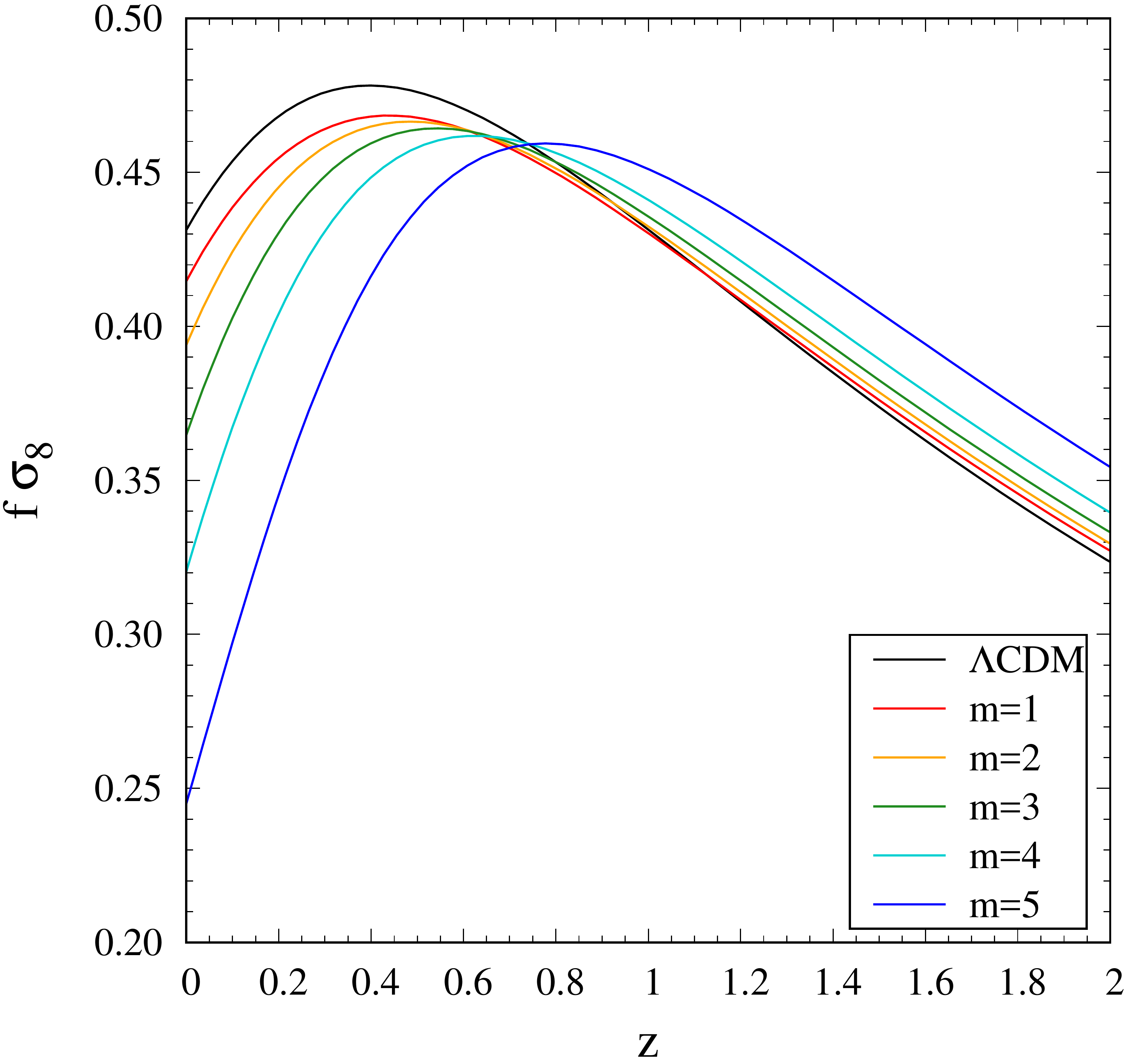}
\end{center}
\caption{\label{fig2}
Evolution of $G_{cc}/G$ (left) and $f\sigma_8$ (right) 
for $\lambda=1$ and $\beta=1/4$ with five different values of $m$. 
The black lines correspond to the case of $\Lambda$CDM model. 
The background initial conditions are the same as those used 
in the simulation of Fig.~\ref{fig1}.
For perturbations, we consider the wavenumber $k=375a_0H_0$ 
with $\sigma_8(z=0)=0.811$. }
\end{figure}

In the left panel of Fig.~\ref{fig2}, we plot the evolution of $G_{cc}/G$ 
for $\lambda=1$ and $\beta=1/4$ with five different values of $m$. 
Since the background dynamics is independent of $m$, 
the five cases shown in Fig.~\ref{fig2} give the same background evolution.
For $\lambda$ and $\beta$ chosen in the simulation of Fig.~\ref{fig2}
the quantity (\ref{rfes}) reduces to 
$r_f=3x_1^2\Omega_c^{-1} m/(6-m)$, 
where $x_1$ and $\Omega_c$ are solely determined by 
the background. As the power $m$ increases in the range 
$0<m<6$, $r_f$ gets larger and hence $G_{cc}$ and 
$G_{cb}$ decrease. Indeed, this property can be confirmed 
in the numerical simulation of Fig.~\ref{fig2}. 
If $m=0$, then the coupling (\ref{fun}) does not contain the 
$Z$ dependence, in which case $G_{cc}=G_{cb}=G$. 
The $Z$ dependence in $f$ allows the possibility for realizing 
the CDM gravitational coupling smaller than $G$. 
When $m=2$, which corresponds to the orange line in Fig.~\ref{fig2},
our interacting model reduces to the scenario 
studied in Ref.~\cite{Pourtsidou:2016ico}. 
Even though the models with same $\beta$ and different $m$ 
give the same background evolution, the dynamics of perturbations 
strongly depends on $m$.
In Fig.~\ref{fig2}, we observe that, for $m>2$, $G_{cc}$ is smaller 
than in the case $m=2$ at low redshifts.
Since the quantity (\ref{rfes}) diverges in the limit 
$m \to 2+1/\beta$, the largest effect for reducing $G_{cc}$ 
and $G_{cb}$ occurs around $m=2+1/\beta$. 

{}From Eqs.~(\ref{delceq2}) and (\ref{delbeq2}), the density contrasts 
of CDM and baryons obey 
\ba
& &
\delta_{c{\rm N}}''+\left( \frac{1}{2}-\frac{3}{2} w_{\rm eff}
+c \right) \delta_{c{\rm N}}'
-\frac{3}{2} \frac{1}{1+r_f} \left( \Omega_c \delta_{c{\rm N}}
+\Omega_b \delta_{b{\rm N}} \right)=0\,,\label{delcev}\\
& &
\delta_{b{\rm N}}''+\left( \frac{1}{2}-\frac{3}{2} w_{\rm eff}
\right) \delta_{b{\rm N}}'
-\frac{3}{2} \left( \Omega_c \delta_{c{\rm N}}
+\Omega_b \delta_{b{\rm N}} \right)=0\,,
\label{delbev}
\ea
where 
\be
c=\frac{2\sqrt{6} \lambda m \beta x_1 x_2^2-6m \beta 
(1+2\beta)x_1^2}{\Omega_c [1+\beta (2-m)]+2m \beta 
(1+2\beta)x_1^2}\,.
\ee
During the deep matter era in which the scalar-field densities are suppressed 
relative to the CDM density ($x_1^2 \ll \Omega_c, x_2^2 \ll \Omega_c$), 
the quantities $r_f$ and $c$ are much smaller than 1. 
In this regime, both $\delta_{c{\rm N}}$ and $\delta_{b{\rm N}}$
grow according to 
$\delta_{c{\rm N}} \propto \delta_{b{\rm N}} \propto a$. 
After the DE density dominates over the CDM density, 
the deviations of $r_f$ and $c$ from 0 tend to be significant.
On the fixed point (\ref{Pdef}), we have $G_{cc}=G_{cb} \to 0$ and 
$c \to 3-\lambda^2/(1+2\beta)$. 
Under the condition (\ref{lamcon}), this asymptotic value of $c$ is 
larger than 1. Numerically, we confirmed that the coefficient 
$1/2-3w_{\rm eff}/2+c$ in front of the friction term $\delta_{c{\rm N}}'$
in Eq.~(\ref{delcev}) remains positive during the transition from 
the matter era to the DE dominance. 

For $r_f>0$ the CDM gravitational coupling $G/(1+r_f)$ decreases 
from $G$ at  low redshifts, so the growth rate of 
$\delta_{c{\rm N}}$ is smaller than that for $r_f=0$. 
The gravitational coupling of baryons is equivalent to $G$, 
but $\delta_{b{\rm N}}$ is affected by the CDM perturbation 
through the quantity $\Omega_c \delta_{c{\rm N}}$ in 
Eq.~(\ref{delbev}). Defining the total matter perturbation as
$\delta \rho_{m{\rm N}}=\delta \rho_{c{\rm N}}+\delta \rho_{b{\rm N}}$ 
with the background density $\rho_m=\rho_c+\rho_b$, the corresponding 
density contrast is 
\be
\delta_{m{\rm N}} \equiv \frac{\delta \rho_{m{\rm N}}}{\rho_m}
=\frac{\Omega_c}{\Omega_m} \delta_{c{\rm N}}
+\frac{\Omega_b}{\Omega_m} \delta_{b{\rm N}}\,,
\label{delm}
\ee
where $\Omega_m=\Omega_c+\Omega_b$. 
The growth rate of matter perturbations is given by 
$f \equiv \dot{\delta}_{m{\rm N}}/(H \delta_{m{\rm N}})
=\delta_{m{\rm N}}'/\delta_{m{\rm N}}$. 
Today's values of CDM and baryon density parameters are 
$\Omega_c \simeq 0.27$ and $\Omega_b \simeq 0.05$, 
respectively, so the dominant contribution to $\delta_{m{\rm N}}$ 
comes from the CDM perturbation.

In the right panel of Fig.~\ref{fig2}, we plot the evolution of 
$f \sigma_8$ for five different cases, where the 
Planck2018 best-fit value $\sigma_8 (z=0)=0.811$ \cite{Aghanim:2018eyx} 
is used in the simulation. 
For increasing $m$ the values of $f\sigma_8$ 
at low redshifts decrease, by reflecting the fact 
that $G_{cc}$ and $G_{cb}$ get smaller. 
In comparison to the $\Lambda$CDM model, our interacting 
model allows the possibility for alleviating the problem of 
$\sigma_8$ tension by reducing the values of $f\sigma_8$. 
Moreover, the model is sufficiently versatile in that almost any cosmic 
growth rate weaker than that in the $\Lambda$CDM model can be realized by 
appropriately choosing the values of $m$ and $\beta$. 

In order to examine whether our interacting model is observationally 
favored over the $\Lambda$CDM model, we need to carry out the 
Markov-chain-Monte-Carlo simulation by varying $\sigma_8 (z=0)$ 
besides other cosmological parameters instead of setting them to 
the Planck best-fit values of the $\Lambda$CDM.
In this vein, Ref.~\cite{Pourtsidou:2016ico} studied observational 
constraints on the interacting DE and DM scenario with $f=\beta Z^2$, 
i.e., the power $m=2$. The joint analysis using CMB and galaxy 
clustering data showed that this particular model is favored over 
the $\Lambda$CDM model as a consequence of alleviating the 
$\sigma_8$ tension. Since the coupling (\ref{fun}) can realize almost 
any cosmic growth rate with the additional power $m$, 
it is of interest to explore how the latest data 
put observational constraints on our model. 
This interesting issue is left for a future work.

{}From Eqs.~(\ref{Psi}) and (\ref{delm}), the gravitational potentials 
are expressed as 
\be
\Psi=-\Phi \simeq -\frac{3}{2} \left( \frac{aH}{k} \right)^2 
\Omega_m \delta_{m{\rm N}}\,. 
\ee
When $r_f>0$ the growth rate of $\delta_{m{\rm N}}$ is smaller 
than that for $r_f=0$, so the gravitational potentials in the former 
decay faster than those in the latter. 
Indeed, this property is confirmed in our numerical simulation. 
The suppressed gravitational potentials at low redshifts, together 
with the absence of gravitational slip, are key features 
for probing our model further from the 
observations of weak lensing.

\section{Conclusions}
\label{consec}

We studied the cosmology of coupled DE and DM theories given by 
the action (\ref{action}). The scalar derivative interaction with the 
CDM four velocity, which is weighed by the scalar combination 
$Z=u_c^{\mu} \partial_{\mu} \phi$, gives rise to several 
different features in comparison to the scalar field coupled 
to the CDM density of the form $f_1(\phi) \rho_c$. 
At the background level the coupling $f(X, Z)$ modifies 
the DE density and pressure in the forms (\ref{rhode}) 
and (\ref{Pde}), but the interacting terms do not explicitly 
appear on the right hands of DE and CDM continuity equations.
This is attributed to the fact that the interaction between DE 
and DM in our scenario corresponds to the momentum 
transfer \cite{Pourtsidou:2013nha,Pourtsidou:2016ico}.

In Sec.~\ref{persec}, we expanded the action (\ref{action}) up to 
second order in scalar perturbations and derived the quadratic-order 
action in the form (\ref{saction}) with the Lagrangians 
(\ref{L0}) and (\ref{Lf}).
We then obtained the full linear perturbation equations of motion 
in the gauge-ready form. 
The gauge-invariant gravitational potentials $\Psi$ and $\Phi$ 
obey the relation (\ref{aniso}), so there is no gravitational slip 
in our coupled DE and DM theories.

In Sec.~\ref{stasec}, we derived conditions for the absence of scalar ghosts 
and Laplacian instabilities by eliminating nondynamical perturbations from 
the second-order action (\ref{saction}).  
In the small-scale limit, the no-ghost conditions are given by 
Eqs.~(\ref{noghost1}) and (\ref{noghost2}).
We showed that, provided $c_c^2=0$, the effective CDM sound speed 
squared $c_{\rm CDM}^2$ vanishes, so there is no additional pressure 
preventing the growth of CDM density perturbations. 
The Laplacian instability of the scalar degree of freedom is absent under 
the condition (\ref{cscon}), where $c_s^2$ is given by Eq.~(\ref{csf}) 
with Eq.~(\ref{css}).
Applying the quasi-static approximation to the perturbations relevant to the growth 
of large-scale structures, we obtained the effective gravitational couplings 
felt by CDM and baryons in the forms (\ref{Gcc}) and (\ref{Gbc}), respectively. 
As long as the quantity $r_f$ defined by Eq.~(\ref{rf}) is positive, 
the CDM gravitational couplings $G_{cc}$ and $G_{cb}$ are smaller 
than the Newton constant $G$.

In Sec.~\ref{modelsec}, we studied the late-time cosmological dynamics 
for the explicit DE and DM coupling of the form (\ref{fun}).
At the background level, the effect of interactions appears only through the 
derivative term $\beta \dot{\phi}^2$ in $\rho_{\rm DE}$ and $P_{\rm DE}$.
For the exponential potential $V(\phi)=V_0 e^{-\lambda \phi/M_{\rm pl}}$, 
the positive coupling $\beta$ leads to the dark energy equation of state closer to 
$-1$ relative to the uncoupled case.
Provided that $1+\beta (2-m)> 0$, the stability conditions (\ref{qscon})-(\ref{cscon2}) 
of scalar perturbations are satisfied for positive values of $\beta$ and $m$.
Since $r_f>0$ in this case, the gravitational interaction with 
CDM is weaker than that for $\beta=0$. 
This property manifests itself in the suppressed growth 
of the total matter density contrast (\ref{delm}).  
For given values of $\beta$ and $\lambda$,
$f \sigma_8$ tends to be smaller with the increase of $m$.
As we observe in Fig.~\ref{fig2}, this allows the suppressed growth 
rate of $\delta_{m{\rm N}}$ at low redshifts in comparison 
to the $\Lambda$CDM model.

We have thus shown that the dark-sector interaction $f(X,Z)$  
provides an interesting possibility for realizing the cosmic growth rate 
weaker than that in the $\Lambda$CDM model. 
In comparison to the model $m=2$ studied in the literature, our 
interacting scenario with arbitrary powers of $m$ gives rise 
to a wide variety of perturbation dynamics which can be 
tested with observations.
The coupling (\ref{fun}) may alleviate the tension of $\sigma_8$ 
between low- and high-redshift measurements by suppressing the values of 
$f\sigma_8$ in the range $z \lesssim 1$. 
It will be of interest to study in detail how much extent the coupling  (\ref{fun}) reduces 
the tensions of $\sigma_8$ and $H_0$ present in the $\Lambda$CDM model. 
We leave observational constraints on our interacting model
for a future work.

\section*{Acknowledgements}

We thank Edmund Copeland for useful correspondence about our 
previous paper \cite{Kase:2019veo} and for motivating us 
to study derivatively coupled DE and DM theories further.
RK is supported by the Grant-in-Aid for Young Scientists B 
of the JSPS No.\,17K14297. ST is supported by the Grant-in-Aid for 
Scientific Research Fund of the JSPS No.\,19K03854 and
MEXT KAKENHI Grant-in-Aid for Scientific Research on Innovative Areas
``Cosmic Acceleration'' (No.\,15H05890).



\begin{thebibliography}{99}

\bibitem{Peebles:1984ge} 
P.~J.~E.~Peebles,
Astrophys.\ J.\  {\bf 284}, 439 (1984).

\bibitem{Peebles:1982ff} 
P.~J.~E.~Peebles,
Astrophys.\ J.\  {\bf 263}, L1 (1982).

\bibitem{Weinberg} 
S.~Weinberg,
Rev.\ Mod.\ Phys.\  {\bf 61}, 1 (1989).

\bibitem{Riess:2016jrr} 
A.~G.~Riess {\it et al.},
Astrophys.\ J.\  {\bf 826}, 56 (2016)
[arXiv:1604.01424 [astro-ph.CO]].

\bibitem{Aghanim:2018eyx} 
N.~Aghanim {\it et al.} [Planck Collaboration],
arXiv:1807.06209 [astro-ph.CO].

\bibitem{Verde:2019ivm} 
L.~Verde, T.~Treu and A.~G.~Riess,
Nature Astronomy, {\bf 3},  891-895 (2019)
[arXiv:1907.10625 [astro-ph.CO]].

\bibitem{Riess:2019cxk}
A.~G.~Riess, S.~Casertano, W.~Yuan, L.~M.~Macri and D.~Scolnic,
Astrophys.\ J.\  {\bf 876}, no. 1, 85 (2019)
[arXiv:1903.07603 [astro-ph.CO]].
  
\bibitem{Freedman:2019jwv}
W.~L.~Freedman {\it et al.},
arXiv:1907.05922 [astro-ph.CO].

\bibitem{Wong:2019kwg}
K.~C.~Wong {\it et al.},
arXiv:1907.04869 [astro-ph.CO].

\bibitem{Reid:2019tiq}
M.~J.~Reid, D.~W.~Pesce and A.~G.~Riess,
Astrophys.\ J.\  {\bf 886}, no. 2, L27 (2019)
[arXiv:1908.05625 [astro-ph.GA]].

\bibitem{Shajib:2019toy}
A.~J.~Shajib {\it et al.} [DES Collaboration],
arXiv:1910.06306 [astro-ph.CO].

\bibitem{Macaulay:2013swa} 
E.~Macaulay, I.~K.~Wehus and H.~K.~Eriksen,
Phys.\ Rev.\ Lett.\  {\bf 111}, 161301 (2013)
[arXiv:1303.6583 [astro-ph.CO]].

\bibitem{Nesseris:2017vor} 
S.~Nesseris, G.~Pantazis and L.~Perivolaropoulos,
Phys.\ Rev.\ D {\bf 96}, 023542 (2017)
[arXiv:1703.10538 [astro-ph.CO]].

\bibitem{Hildebrandt:2016iqg} 
H.~Hildebrandt {\it et al.},
Mon.\ Not.\ Roy.\ Astron.\ Soc.\  {\bf 465}, 1454 (2017)
[arXiv:1606.05338 [astro-ph.CO]].

\bibitem{Joudaki:2017zdt} 
S.~Joudaki {\it et al.},
Mon.\ Not.\ Roy.\ Astron.\ Soc.\  {\bf 474}, 4894 (2018)
[arXiv:1707.06627 [astro-ph.CO]].

\bibitem{Wands} 
V.~Salvatelli, N.~Said, M.~Bruni, A.~Melchiorri and D.~Wands,
Phys.\ Rev.\ Lett.\  {\bf 113}, 181301 (2014)
[arXiv:1406.7297 [astro-ph.CO]].

\bibitem{Kumar:2016zpg} 
S.~Kumar and R.~C.~Nunes,
Phys.\ Rev.\ D {\bf 94}, 123511 (2016)
[arXiv:1608.02454 [astro-ph.CO]].

\bibitem{Kumar:2017dnp} 
S.~Kumar and R.~C.~Nunes,
Phys.\ Rev.\ D {\bf 96}, 103511 (2017)
[arXiv:1702.02143 [astro-ph.CO]].

\bibitem{DiValentino:2017iww} 
E.~Di Valentino, A.~Melchiorri and O.~Mena,
Phys.\ Rev.\ D {\bf 96}, 043503 (2017)
[arXiv:1704.08342 [astro-ph.CO]].

\bibitem{An:2017crg} 
R.~An, C.~Feng and B.~Wang,
JCAP {\bf 1802}, 038 (2018)
[arXiv:1711.06799 [astro-ph.CO]].

\bibitem{Kaza} 
L.~Kazantzidis and L.~Perivolaropoulos,
Phys.\ Rev.\ D {\bf 97}, 103503 (2018)
[arXiv:1803.01337 [astro-ph.CO]].

\bibitem{Yang:2018euj} 
W.~Yang, S.~Pan, E.~Di Valentino, R.~C.~Nunes, S.~Vagnozzi and D.~F.~Mota,
JCAP {\bf 1809}, 019 (2018)
[arXiv:1805.08252 [astro-ph.CO]].

\bibitem{Pan:2019gop} 
S.~Pan, W.~Yang, E.~Di Valentino, E.~N.~Saridakis and S.~Chakraborty,
arXiv:1907.07540 [astro-ph.CO].

\bibitem{DiValentino:2019ffd} 
E.~Di Valentino, A.~Melchiorri, O.~Mena and S.~Vagnozzi,
arXiv:1908.04281 [astro-ph.CO].

\bibitem{Yang:2019uog} 
W.~Yang, S.~Pan, R.~C.~Nunes and D.~F.~Mota,
arXiv:1910.08821 [astro-ph.CO].

\bibitem{Dalal:2001dt} 
N.~Dalal, K.~Abazajian, E.~E.~Jenkins and A.~V.~Manohar,
Phys.\ Rev.\ Lett.\  {\bf 87}, 141302 (2001)
[astro-ph/0105317].

\bibitem{Zimdahl:2001ar} 
W.~Zimdahl and D.~Pavon,
Phys.\ Lett.\ B {\bf 521}, 133 (2001)
[astro-ph/0105479].

\bibitem{Chimento:2003iea} 
L.~P.~Chimento, A.~S.~Jakubi, D.~Pavon and W.~Zimdahl,
Phys.\ Rev.\ D {\bf 67}, 083513 (2003)
[astro-ph/0303145].

\bibitem{Wang1} 
B.~Wang, Y.~g.~Gong and E.~Abdalla,
Phys.\ Lett.\ B {\bf 624}, 141 (2005)
[hep-th/0506069].

\bibitem{Wei:2006ut} 
H.~Wei and S.~N.~Zhang,
Phys.\ Lett.\ B {\bf 644}, 7 (2007)
[astro-ph/0609597].

\bibitem{Amendola:2006dg} 
L.~Amendola, G.~Camargo Campos and R.~Rosenfeld,
Phys.\ Rev.\ D {\bf 75}, 083506 (2007)
[astro-ph/0610806].

\bibitem{Guo:2007zk} 
Z.~K.~Guo, N.~Ohta and S.~Tsujikawa,
Phys.\ Rev.\ D {\bf 76}, 023508 (2007)
[astro-ph/0702015 [astro-ph]].

\bibitem{Gavela:2009cy} 
M.~B.~Gavela, D.~Hernandez, L.~Lopez Honorez, O.~Mena and S.~Rigolin,
JCAP {\bf 0907}, 034 (2009)
[arXiv:0901.1611 [astro-ph.CO]].

\bibitem{Jackson:2009mz} 
B.~M.~Jackson, A.~Taylor and A.~Berera,
Phys.\ Rev.\ D {\bf 79}, 043526 (2009)
[arXiv:0901.3272 [astro-ph.CO]].

\bibitem{Faraoni:2014vra} 
V.~Faraoni, J.~B.~Dent and E.~N.~Saridakis,
Phys.\ Rev.\ D {\bf 90}, 063510 (2014)
[arXiv:1405.7288 [gr-qc]].

\bibitem{Tamanini:2015iia} 
N.~Tamanini,
Phys.\ Rev.\ D {\bf 92}, 043524 (2015)
[arXiv:1504.07397 [gr-qc]].

\bibitem{Valiviita:2008iv} 
J.~Valiviita, E.~Majerotto and R.~Maartens,
JCAP {\bf 0807}, 020 (2008)
[arXiv:0804.0232 [astro-ph]].

\bibitem{Valiviita:2009nu} 
J.~Valiviita, R.~Maartens and E.~Majerotto,
Mon.\ Not.\ Roy.\ Astron.\ Soc.\  {\bf 402}, 2355 (2010)
[arXiv:0907.4987 [astro-ph.CO]].

\bibitem{Pourtsidou:2013nha} 
A.~Pourtsidou, C.~Skordis and E.~J.~Copeland,
Phys.\ Rev.\ D {\bf 88}, 083505 (2013)
[arXiv:1307.0458 [astro-ph.CO]].

\bibitem{Boehmer:2015kta} 
C.~G.~Boehmer, N.~Tamanini and M.~Wright,
Phys.\ Rev.\ D {\bf 91}, 123002 (2015)
[arXiv:1501.06540 [gr-qc]].

\bibitem{Boehmer:2015sha} 
C.~G.~Boehmer, N.~Tamanini and M.~Wright,
Phys.\ Rev.\ D {\bf 91}, 123003 (2015)
[arXiv:1502.04030 [gr-qc]].

\bibitem{Skordis:2015yra} 
C.~Skordis, A.~Pourtsidou and E.~J.~Copeland,
Phys.\ Rev.\ D {\bf 91}, 083537 (2015)
[arXiv:1502.07297 [astro-ph.CO]].

\bibitem{Koivisto:2015qua} 
T.~S.~Koivisto, E.~N.~Saridakis and N.~Tamanini,
JCAP {\bf 1509}, 047 (2015)
[arXiv:1505.07556 [astro-ph.CO]].

\bibitem{Pourtsidou:2016ico} 
A.~Pourtsidou and T.~Tram,
Phys.\ Rev.\ D {\bf 94}, 043518 (2016)
[arXiv:1604.04222 [astro-ph.CO]].

\bibitem{Dutta:2017kch} 
J.~Dutta, W.~Khyllep and N.~Tamanini,
Phys.\ Rev.\ D {\bf 95}, 023515 (2017)
[arXiv:1701.00744 [gr-qc]].

\bibitem{Kase:2019veo} 
R.~Kase and S.~Tsujikawa,
Phys.\ Rev.\ D {\bf 101}, 063511 (2020)
[arXiv:1910.02699 [gr-qc]].

\bibitem{Sorkin}
B.~F.~Schutz and R.~Sorkin,
Annals Phys.\  {\bf 107}, 1 (1977).

\bibitem{Brown} 
J.~D.~Brown,
Class.\ Quant.\ Grav.\  {\bf 10}, 1579 (1993)
[gr-qc/9304026].

\bibitem{Bettoni:2011fs} 
D.~Bettoni, S.~Liberati and L.~Sindoni,
JCAP {\bf 1111}, 007 (2011)
[arXiv:1108.1728 [gr-qc]].

\bibitem{Bettoni:2015wla} 
D.~Bettoni and S.~Liberati,
JCAP {\bf 1508}, 023 (2015)
[arXiv:1502.06613 [gr-qc]].

\bibitem{DGS}
A.~De Felice, J.~M.~Gerard and T.~Suyama,
Phys.\ Rev.\ D {\bf 81}, 063527 (2010)
[arXiv:0908.3439 [gr-qc]].

\bibitem{DeFelice:2016yws} 
A.~De Felice, L.~Heisenberg, R.~Kase, S.~Mukohyama, S.~Tsujikawa and Y.~l.~Zhang,
JCAP {\bf 1606}, 048 (2016)
[arXiv:1603.05806 [gr-qc]].

\bibitem{Kase:2018nwt} 
R.~Kase and S.~Tsujikawa,
JCAP {\bf 1811}, 024 (2018)
[arXiv:1805.11919 [gr-qc]].

\bibitem{Frusciante:2018tvu} 
N.~Frusciante, R.~Kase, K.~Koyama, S.~Tsujikawa and D.~Vernieri,
Phys.\ Lett.\ B {\bf 790}, 167 (2019)
[arXiv:1812.05204 [gr-qc]].

\bibitem{Naka2019} 
S.~Nakamura, R.~Kase and S.~Tsujikawa,
JCAP {\bf 1912}, 032 (2019)
[arXiv:1907.12216 [gr-qc]].

\bibitem{Barros} 
B.~J.~Barros,
Phys.\ Rev.\ D {\bf 99}, 064051 (2019)
[arXiv:1901.03972 [gr-qc]].

\bibitem{Amendola99a} 
L.~Amendola,
Phys.\ Rev.\ D {\bf 60}, 043501 (1999)
[astro-ph/9904120].

\bibitem{Fujii} 
Y. Fujii and K.~Maeda, 
``The scalar-tensor theory of gravitation'', 
Cambridge University Press (2003).

\bibitem{Wetterich} 
C.~Wetterich,
Astron.\ Astrophys.\  {\bf 301}, 321 (1995)
[hep-th/9408025].

\bibitem{Amendola99} 
L.~Amendola,
Phys.\ Rev.\ D {\bf 62}, 043511 (2000)
[astro-ph/9908023].

\bibitem{Gumjudpai} 
B.~Gumjudpai, T.~Naskar, M.~Sami and S.~Tsujikawa,
JCAP {\bf 0506}, 007 (2005)
[hep-th/0502191].

\bibitem{Amendola06} 
L.~Amendola, M.~Quartin, S.~Tsujikawa and I.~Waga,
Phys.\ Rev.\ D {\bf 74}, 023525 (2006)
[astro-ph/0605488].

\bibitem{Bardeen} 
J.~M.~Bardeen,
Phys.\ Rev.\ D {\bf 22}, 1882 (1980).

\bibitem{Hwang:2001qk} 
J.~c.~Hwang and H.~r.~Noh,
Phys.\ Rev.\ D {\bf 65}, 023512 (2002)
[astro-ph/0102005].

\bibitem{Heisenberg:2018wye} 
L.~Heisenberg, R.~Kase and S.~Tsujikawa,
Phys.\ Rev.\ D {\bf 98}, 123504 (2018)
[arXiv:1807.07202 [gr-qc]].

\bibitem{Heisenberg:2018yae} 
L.~Heisenberg, M.~Bartelmann, R.~Brandenberger and A.~Refregier,
Phys.\ Rev.\ D {\bf 98}, 123502 (2018)
[arXiv:1808.02877 [astro-ph.CO]].

\bibitem{Akrami} 
Y.~Akrami, R.~Kallosh, A.~Linde and V.~Vardanyan,
Fortsch.\ Phys.\  {\bf 67}, no. 1-2, 1800075 (2019)
[arXiv:1808.09440 [hep-th]].


\end{thebibliography}
\end{document}